\documentclass[superscriptaddress,letterpaper,aps,twocolumn]{revtex4-1}

\usepackage{graphicx}
\usepackage{amssymb,amsfonts,amsmath,xcolor}
\usepackage{setspace}

\newcommand{\mb}[1]{\mathbf{#1}}

\newcommand{\hmb}[1]{\hat{\mathbf{#1}}}
\newcommand{\nocontentsline}[3]{}
\newcommand{\tocless}[2]{\bgroup\let\addcontentsline=\nocontentsline#1{#2}\egroup}

\DeclareRobustCommand{\gobblefour}[4]{}
\newcommand*{\SkipTocEntry}{\addtocontents{toc}{\gobblefour}}

\DeclareRobustCommand{\gobbleone}[4]{}

\begin{document}

\title{Design principles governing the motility of myosin V}

\author{Michael Hinczewski} 
\affiliation{Biophysics Program, Institute for Physical Science and
  Technology, University of Maryland, College Park, MD 20742, USA}

\email{mhincz@umd.edu or thirum@umd.edu} 

\author{Riina Tehver}
\affiliation{Department of Physics, Denison University, Granville, OH 43023, USA}

\author{D. Thirumalai}
\affiliation{Biophysics Program, Institute for Physical Science and
  Technology, University of Maryland, College Park, MD 20742, USA}

\begin{abstract}
  The molecular motor myosin V exhibits a wide repertoire of pathways
  during the stepping process, which is intimately connected to its
  biological function.  The best understood of these is the
  hand-over-hand stepping by a swinging lever arm movement toward the
  plus-end of actin filaments, essential to its role as a cellular
  transporter.  However, single-molecule experiments have also shown
  that the motor ``foot stomps'', with one hand detaching and
  rebinding to the same site, and backsteps under sufficient
  load. Explaining the complete taxonomy of myosin V's load-dependent
  stepping pathways, and the extent to which these are constrained by
  motor structure and mechanochemistry, are still open questions.
  Starting from a polymer model, we develop an analytical theory to
  understand the minimal physical properties that govern motor
  dynamics.  In particular, we solve the first-passage problem of the
  head reaching the target binding site, investigating the competing
  effects of load pulling back at the motor, strain in the leading
  head that biases the diffusion in the direction of the target, and
  the possibility of preferential binding to the forward site due to
  the recovery stroke.  The theory reproduces a variety of
  experimental data, including the power stroke and slow diffusive
  search regimes in the mean trajectory of the detached head, and the
  force dependence of the forward-to-backward step ratio, run length,
  and velocity.  The analytical approach yields a formula for the
  stall force, identifying the relative contributions of the chemical
  cycle rates and mechanical features like the bending rigidities
  of the lever arms.  Most importantly, by fully exploring the design space of
  the motor, we predict that myosin V is a robust motor whose
  dynamical behavior is not compromised by reasonable perturbations to
  the reaction cycle, and changes in the architecture of the lever
  arm.
  \end{abstract}

\maketitle

\begin{figure*}
\centerline{\includegraphics[width=\textwidth]{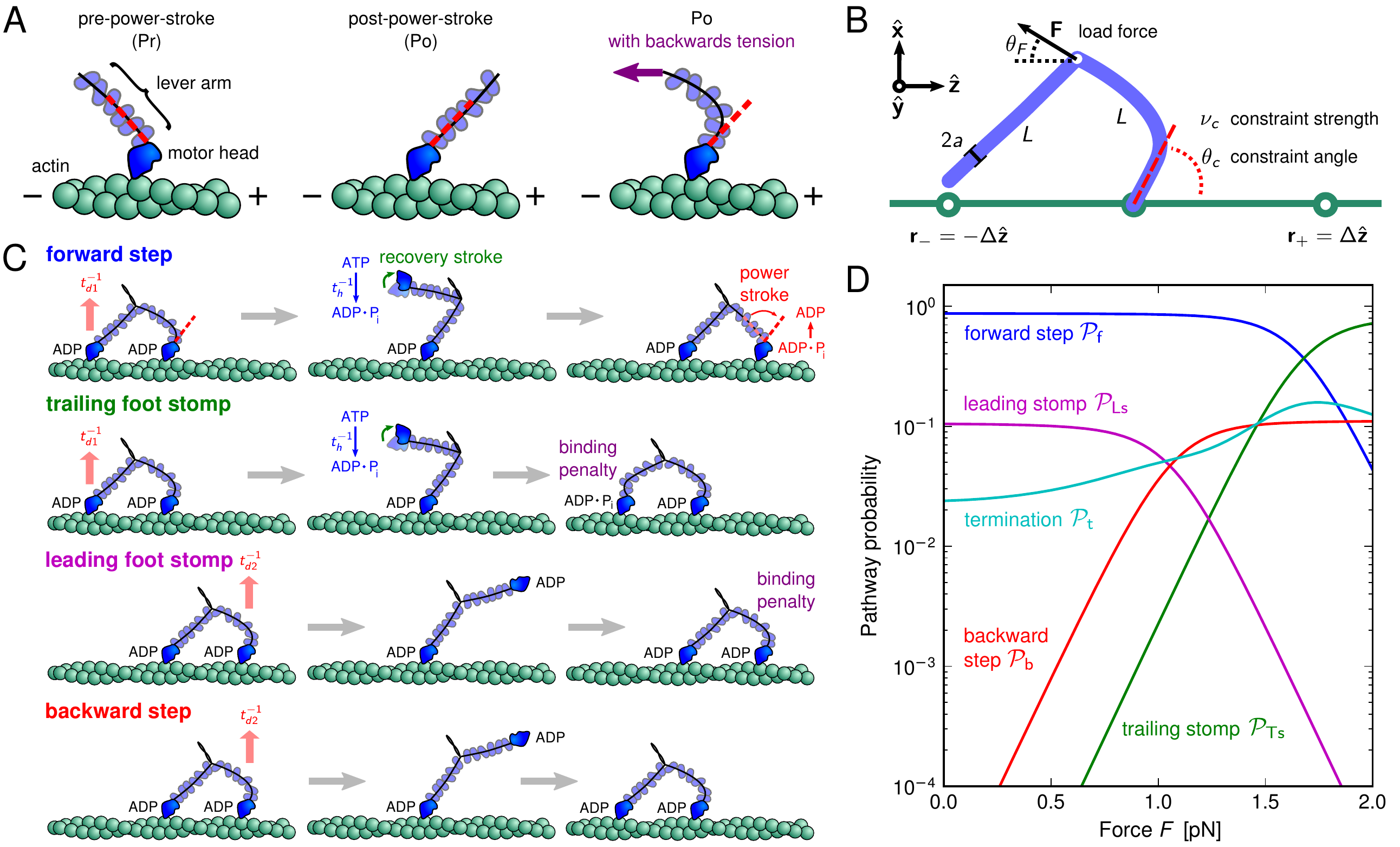}}
\caption{A) Orientational states of the MyoV head with respect to its
  lever arm.  For clarity, only one leg of the two-legged motor is
  shown, though the states are the same for both legs.  Left:
  pre-power-stroke (Pr); center: post-power-stroke (Po).  For each
  state the relaxed orientation (in the absence of tension on the
  lever arm end) is marked by a dashed red line.  Right: the Po state
  with backward tension on the arm end, causing the lever arm to bends
  backwards away from its relaxed direction.  B) The coarse-grained
  polymer representation of MyoV.  C) Schematic view of four MyoV
  kinetic pathways.  For simplicity, the apo state, following ADP
  release from the trailing head and before ATP binding, is not shown.
  D) The probability of each kinetic pathway as a function of backward
  force $F$ (with $\theta_F=0$) calculated from the theory using the
  parameter set in Table~1.}\label{path}
\end{figure*}

  Myosin V (MyoV), a cytoskeletal motor protein belonging to the
  myosin superfamily~\cite{Sivaramakrishnan07}, converts energy from
  ATP hydrolysis into the transport of intracellular cargo such as
  mRNA and organelles along actin filaments~\cite{Reck-Peterson00}.
  In its dimeric form the motor has two actin-binding, ATPase heads,
  connected to alpha-helical lever arm domains stiffened by attached
  calmodulins or essential light chains (Fig.~\ref{path}).  The
  nucleotide-driven mechanochemical cycle of the heads produces two
  changes in the lever arm orientation: a power stroke, where an
  actin-bound head swings the lever arm forward toward the plus
  (barbed) end of the filament, and a recovery stroke which returns
  the arm to its original configuration when the head is detached from
  actin~\cite{Shiroguchi11}.  The motor translates these changes into
  processive plus-end-directed
  movement~\cite{Mehta99,Rief00,Sakamoto00}.  By alternating head
  detachment, MyoV walks hand-over-hand~\cite{Yildiz03,Forkey03},
  taking one $\approx 36$ nm step for each ATP
  consumed~\cite{Sakamoto08}.  At small loads the motor can complete
  $\approx 20-60$ forward steps before dissociating from
  actin~\cite{Sakamoto00,Baker04, Pierobon09}.  Such a high
  unidirectional processivity requires coordination in the detachment
  of the two heads, a ``gating'' mechanism which is believed to arise
  from the strain within the molecule when both heads are bound to
  actin~\cite{Veigel02,Rosenfeld04,Veigel05,Purcell05}.  Sufficiently
  large opposing loads can counteract the plus-end-directed bias,
  resulting in an increase in the probability of
  backstepping~\cite{Kad08} until the motor velocity goes to zero at a
  stall force $\approx 1.9-3$
  pN~\cite{Mehta99,Veigel02,Uemura04,Gebhardt06,Cappello07,Kad08}.
  Although MyoV is among the most extensively studied of motor
  proteins, improvements in experimental resolution continue to
  provide new and surprising insights into the details of its
  dynamics.  A beautiful recent example is the high-speed atomic force
  microscopy (AFM) of Kodera {\em et. al.}~\cite{Kodera10}, which was
  used to visualize not only the expected hand-over-hand stepping, but
  additional, less well-understood processes like ``foot
  stomping''~\cite{Syed06,Beausang13}, where one head detaches and
  rebinds to the same site.  Thus, a comprehensive picture of MyoV
  motility needs to account for all the kinetic pathways, including
  backstepping and foot stomping, how they vary under load, and their
  relationship to the structural and chemical parameters of the motor.

  To address these issues, we introduce a minimal model of MyoV
  dynamics, focusing on the stochastic fluctuations of the motor head
  during the diffusive search of the detached head for a binding site,
  whose importance has been illuminated by various
  experiments~\cite{Dunn07,Shiroguchi07,Komori07,Beausang13}.  The
  large persistence length of the lever
  arms~\cite{Howard96,Moore04,Vilfan05} allows us to propose a novel
  coarse-grained polymer model for the reaction-diffusion problem,
  which in turn yields approximate analytical expressions for all the
  physical observables, including binding times, run length, velocity,
  and stall force.  We have built on the insights of earlier
  theoretical
  works~\cite{Kolomeisky03,Vilfan05,Lan05,Skau06,Tsygankov07,Xu09,Bierbaum11},
  which focused on modeling a reaction network of discrete states in
  the mechanochemical cycle of the motor heads. Our work supplements
  the reaction network with an explicit treatment of the diffusive
  search, which has been studied using insightful Brownian dynamics
  simulations of forward stepping in MyoV~\cite{Craig09}.  An
  important aspect of our theory is that it allows us to tackle not
  just forward steps but the full complexity of foot stomping and
  backstepping across the entire force spectrum up to the stall point.
  In our framework, the load dependence of the MyoV behavior enters
  naturally, as pulling on the molecule shifts the speed and
  likelihood of the detached head reaching the forward or backward
  binding sites.  The competition between the time scales of first
  passage to the sites, and how they compare to the detachment rates
  of the heads, determines the partitioning of the kinetic pathways.
  Significantly, polymer theory gives us a direct connection between
  the kinetics and the structural features of the motor, like the
  bending elasticity of the lever arms and the orientational bias due
  to the power stroke.  The result is a theory with only three fitting
  parameters that have not been previously determined through
  experiment, all of which have simple physical interpretations.  The
  theoretical fit quantitatively reproduces a variety of experimental data, like the
  time-dependent mean trajectories of the detached head~\cite{Dunn07},
  and the force dependence of the backward-to-forward step
  ratio~\cite{Kad08} and run length /
  velocity~\cite{Mehta99,Clemen05,Uemura04,Gebhardt06,Kad08}.  We also
  explore more broadly the design space of MyoV structural parameters,
  allowing us to predict the essential requirements for the observed
  dynamical behavior, and to answer the following questions.  Is the
  structure of the motor dictated by certain natural constraints?  How
  robust is the motility of MyoV to perturbations in the parameters?
  What are the relative contributions of head chemistry (resulting
  from changes in the nucleotide states) and the structural features
  to the measured stall force?  The answers to these questions, which
  are provided in terms of phase diagrams, lead to testable
  predictions.

\tocless\section{Results}

\tocless\subsection{Polymer model for MyoV}  In our model for MyoV
(Fig.~\ref{path}B), the motor and lever arm domains of each head are
represented as a single semiflexible polymer chain with contour length
$L$ and persistence length $l_p$.  The two polymer legs are connected
at a freely rotating joint.  The parameter values characterizing our
model are listed in Table~1.  Though the tail domain of
MyoV, attached to the cargo, is not explicitly included, its effect is
to transmit a load force $\mb{F}$ to the joint.  The force is oriented
in the $\hmb{x}-\hmb{z}$ plane, at an angle $\theta_F$, measured
clockwise from $-\hmb{z}$.  The axis $\hmb{z}$ runs parallel to the
actin filament, pointing toward the plus end.  Our focus here is to
study backwards or resistive load ($0 \le \theta_F < 90^\circ$) at
force magnitudes smaller or close to the stall, $F \lesssim
F_\text{stall} \approx 1.9-3$
pN~\cite{Mehta99,Veigel02,Uemura04,Gebhardt06,Cappello07,Kad08}.  The
polymer end-points can bind to the actin filament at discrete binding
sites, which are evenly spaced at a distance $\Delta = 36$ nm along
the filament, corresponding approximately to the half-pitch of the actin
double-helical structure (13 G-actin subunits).  Though the model can
be extended to incorporate a distribution of $\Delta$ values,
reflecting binding to subunits neighboring the primary binding sites,
in the simplest approximation we keep $\Delta$ fixed.  Since the the
first passage times to the primary binding site and its neighbors are
similar, the effect of this approximation is small.

\begin{table}
{\small
\centering
\begin{tabular*}{\columnwidth}{lcc}
\hline
Parameter & Value & Notes\\
\hline
\multicolumn{3}{c}{{\em Mechanical parameters}}\\
\hline
leg contour length $L$ & 35 nm & [35]\\
leg persistence length $l_p$ & 310 nm & [27]\\
head diffusivity $D_\text{h}$  & $5.7\times 10^{-7}$ cm$^2$/s & [40, 41]\\
constraint angle $\theta_c$ & $60^\circ$ & FTE~[23]\\
constraint strength $\nu_c$ & 184 & FTE~[16]\\
\hline
\multicolumn{3}{c}{{\em Binding parameters}}\\
\hline
binding site separation $\Delta$  & 36 nm & [35]\\
capture radius $a$ & 1 nm&\\
binding penalty $b$ & 0.065 & FTE~[6, 10, 11]\\
\hline
\multicolumn{3}{c}{{\em Chemical rates}}\\
\hline
hydrolysis rate $t_\text{h}^{-1}$ & 750 s$^{-1}$& [38]\\
TH detachment rate $t_\text{d1}^{-1}$ & 12 s$^{-1}$& [38]\\
LH detachment rate $t_\text{d2}^{-1}$ & 1.5 s$^{-1}$& [15]\\
gating ratio $g=t_\text{d2}/t_\text{d1}$ & 8&\\
\end{tabular*}}
\caption{Parameters in the model for MyoV dynamics.  FTE denotes a parameter derived from a fit to experimental data.}\label{param}
\end{table}

For each leg, the lever arm can adopt different preferred
configurations with respect to the motor head during the course of the
stepping cycle: the pre-power-stroke (Pr) and post-power-stroke (Po)
states.  When the motor head is bound to actin, and there is no
tension on the end of the lever arm transmitted through the junction,
the two states have relaxed configurations illustrated in
Fig.~\ref{path}A (left, center).  In the Pr state, the lever arm
relaxes to an orientation tilting toward the actin minus end, while in
the Po state it tilts toward the plus end.  In our model, the tilting
preference of the Po state enters as a harmonic constraint on the
end-tangent of the bound leg: if $\hmb{u}_0$ is the unit tangent
vector at the point where the polymer leg attaches to actin, we have a
potential ${\cal H}_c = \frac{1}{2}k_B T \nu_c (\hat{\mb{u}}_0 -
\hat{\mb{u}}_c)^2$, with a constraint strength $\nu_c$ and direction
$\hat{\mb{u}}_c$.  The vector $\hmb{u}_c$ is in the $\hmb{x}-\hmb{z}$
plane at an angle $0< \theta_c< \pi/2$, measured counter-clockwise
from the $+\hmb{z}$ axis (the $\hmb{u}_c$ direction is marked by a red
dashed line in Fig.~\ref{path}A,B).  In principle, the Pr state is
analogous, but with distinct values of $\nu_c$ and $\theta_c$, with
the latter in the range $\pi/2 <\theta_c<\pi$.  However, as we will
see below, all the kinetic pathways involve diffusion while the bound
leg is in the Po state, so the parameters of the Pr state do not
explicitly enter the calculation.  Hence, both $\nu_c$ and
$\theta_c$ will refer only to the Po state.

If there is tension propagated through the junction on the end of the
lever arm (i.e. due to load, or the fact that both motor heads are
bound to actin), the lever arm contour will be bent away from its
relaxed conformation.  Fig.~\ref{path}A (right) shows the Po state
under backward tension on the arm: the lever arm is bent, adopting a
shape that reflects several competing physical effects. The Po
constraint of strength $\nu_c$ tries to keep the head-arm angle near
$\theta_c$, the bending stiffness $l_p$ favors a straight lever arm
contour, and the tension tries to pull the end of the arm backwards.
The polymer model naturally incorporates the interplay of these
effects, which we will show is crucial in determining the dynamical
response of the motor to load.

\vspace{2em}

\tocless\subsection{Kinetic pathways}  The starting point for all MyoV
kinetic pathways (Fig.~\ref{path}C, left column) is the waiting state,
where both heads have ADP, are strongly bound to actin, and are in the
Po state.  Because the leading (L) leg is connected to the trailing
(T) leg at the junction, the L leg is under backward tension, and it
bends in the manner discussed above.  The resulting strained
``telemark'' or ``reverse arrowhead'' stance has been observed
directly in both electron microscopy~\cite{Walker00} and
AFM~\cite{Kodera10} images.  The waiting state leads to four possible
kinetic pathways (Fig.~\ref{path}C):

{\em 1. Forward step.} ADP is released from the trailing head (TH),
followed by ATP binding, which makes association of the head with
actin weak, leading to detachment.  We assume saturating ATP concentrations
($> 100\:\mu$M), where ATP binding and subsequent TH detachment is very
fast compared with ADP release, and hence the entire detachment
process for the TH is modeled with a single rate $t^{-1}_\text{d1}
= 12$ s$^{-1}$, equal to the experimentally measured ADP release
rate~\cite{DeLaCruz99}.  If we set the origin ($z=0$) at the position
of the bound leading head (LH), the free end of MyoV can diffuse and potentially
rebind at one of two sites, $\mb{r}_{\pm} = \pm \Delta \hmb{z}$ along
the actin filament (Fig.~\ref{path}B).  Binding at $\mb{r}_+$ leads to
a forward step (Fig.~\ref{path}A, row 1).  However successful binding
is dependent on two conditions: (i) reaching the capture radius around
the binding site; (ii) the motor head having already hydrolyzed its
bound ATP.

During the diffusive search, the entire two-legged polymer structure
fluctuates in three dimensions, subject only to the end-tangent
constraint at the bound leg attachment point.  First passage to a
given binding site $\mb{r}_\pm$, which occurs at a mean time interval
$t_\text{fp}^\pm$ after detachment, is the first arrival of the
detached head to any point within a radius $a$ of the binding site.
The capture radius $a$, which reflects the distance at which the free
MyoV head can appreciably interact with the actin binding
site~\cite{Craig09}, is set to $a = 1$ nm, comparable to the Debye
screening length $\lambda_D$ in physiological and {\em in vitro}
conditions (i.e. for KCl concentrations of $25-400$ mM, $\lambda_D
\approx 1.9-0.5$ nm).

The second condition for successful binding is the chemical state of
the detached head.  In order for the head to strongly associate and
bind to actin, ATP must hydrolyze to ADP$+$P$_\text{i}$, which occurs
at a rate $t_\text{h}^{-1} = 750$ s$^{-1}$~\cite{DeLaCruz99}. Along
with hydrolysis, the detached head also undergoes a recovery stroke,
which reverses the power stroke, changing the orientation of the head
with respect to the lever arm (Po $\to$ Pr).  For simplicity, we
combine the nucleotide / head-arm orientation states of the detached
head into two possibilities: A) ATP / Po, B) ADP+P$_\text{i}$ / Pr.
Unless otherwise specified, we assume the transition A$\to$B occurs
irreversibly at a rate $t_h^{-1}$.  (We will discuss one experimental
variant of MyoV with modified light chain composition in the section
on zero load binding kinetics, where there is a non-negligible reverse
hydrolysis rate $t_{-\text{h}}^{-1}$.)  Binding can only occur in
state B, so if the detached TH has reached the capture radius of one
of the sites and the system is still in state A, it has zero
probability of binding, resulting in the TH continuing its diffusive
trajectory.  For forward stepping to occur, the TH must reach the
capture radius of $\mb{r}_+$ in state B, and then it can bind with
probability 1.

After successful binding, P$_\text{i}$ is rapidly released from the
bound head, which then results in a Pr $\to$ Po transition, returning
the motor to its waiting state, with both the heads being in the Po
state.  Release of the inorganic phosphate P$_\text{i}$ and the power
stroke are much faster than the detachment time scale
$t_{d1}$~\cite{Rosenfeld04}, so we can assume that the motor with two
bound heads spends nearly all its time waiting in the telemark stance.

{\em 2. Trailing foot stomp.}  This kinetic pathway (Fig.~\ref{path}C,
row 2) is similar to the forward step, except that the detached TH
diffuses to the site $\mb{r}_-$ rather than $\mb{r}_+$.  Rebinding at
$\mb{r}_-$ brings the center-of-mass of the motor back to its original
location, without any net movement along the actin.  For the binding
to be successful, the head must be in state B within the capture
radius $a$ of $\mb{r}_-$, in which case it will bind with a
probability $b < 1$.  The reduced probability of binding is a crucial
difference between the forward step and T foot stomp pathways.  The
binding penalty arises because the head in state B, after the recovery
stroke, is in the Pr orientation, which is believed to favor binding
to the forward target site ($\mb{r}_+$) over the backward site
($\mb{r}_-$)~\cite{Shiroguchi11}.  Forward binding involves the
detached head going in front of its lever arm, which has to tilt back
towards the actin minus end (the relaxed configuration of the Pr
state).  Backward binding has the opposite arrangement, with the lever
arm bent towards the actin plus end, which is an unnatural
configuration in the Pr state, resulting in a strained back leg, as
illustrated on the right in Fig.~\ref{path}C, row 2.  We model this
effective extra energy barrier in the binding process through the
probability $b$.  The greater the barrier, the smaller the value of
$b$.  The hypothesis that the recovery stroke is important in favoring
forward binding has found support in a recent single molecule study on
single-headed MyoV~\cite{Shiroguchi11}, which established that the Pr
orientation is highly kinetically and energetically stable (with an
energy barrier of at least 5 $k_BT$ with respect to Po).

{\em 3. Leading foot stomp.}  In addition to the two kinetic pathways
above, initiated by TH detachment, there are two other possibilities,
that occur upon detachment of the LH.  The first of these is the
leading foot stomp, where the LH unbinds and then rebinds to its
original site (Fig.~\ref{path}C, row 3).  The detachment of the LH
occurs at a slower rate than TH detachment, $t_{d2}^{-1} = (g
t_\text{d1})^{-1}$, where we denote the factor $g > 1$ as the gating
ratio. This asymmetry arises from the intramolecular strain within the
two-legged MyoV structure bound to
actin~\cite{Veigel02,Rosenfeld04,Veigel05,Purcell05}.  The backward
tension on the L lever arm in the waiting state slows down ADP release
in the LH by 50-70 fold compared to the
TH~\cite{Rosenfeld04,Kodera10}, which makes detachment through the ADP
release / ATP binding mechanism very rare.  Rather, the LH under
backwards strain detaches primarily by an alternate pathway where it
retains ADP~\cite{Purcell05,Kodera10}, an assumption supported by the
observation that single-headed MyoV under backwards loads of $\sim 2$
pN unbinds from actin at a slow rate of 1.5 s$^{-1}$ independent of
both ATP and ADP concentrations~\cite{Purcell05}.  As described below,
the magnitude of the backward tension in the waiting state can also be
directly estimated from the structural parameters of the polymer
model, giving a value of $2.7$ pN, sufficient to be in the
slow unbinding regime.  Based on these considerations, we set
$t_\text{d2}^{-1} = 1.5$ s$^{-1}$ in our model, giving a gating ratio
$g=8$.  In other words, the TH is 8 times more likely to detach than
the LH per unit time.  We also assume the LH always retains ADP upon
detachment (staying in the Po state~\cite{Kodera10}) and thus no ATP
hydrolysis needs to occur before rebinding.

If we assign $z=0$ to be the position of the bound TH,
then the L foot stomp involves reattachment to its original site
$\mb{r}_+$.  Since the LH is Po, rebinding requires the lever arm
to be bent backwards, contrary to the plus-directed relaxed
orientation of the Po state.  We thus have a binding penalty analogous
to the one for the T foot stomp: successful binding will occur with a
probability $b$ within the capture radius $a$ around $\mb{r}_+$.
There is no additional chemical requirement, since the LH is in an
ADP state with high affinity to actin.  Though it is possible to
assign a distinct binding penalty $b$ for the T and L foot stomps,
this does not lead to any major qualitative differences in the
analysis below, so we assume for simplicity a single value of $b$.
After binding, MyoV returns to the waiting state.

{\em 4. Backward step.}  The final kinetic pathway proceeds
analogously to the L foot stomp, but the detached LH diffuses and
binds to the backward site $\mb{r}_-$ (Fig.~\ref{path}C, row 4).  MyoV
thus steps backwards,  shifting the center-of-mass towards the minus
end of actin.  The detached head retains ADP, and stays in the Po
state.  Because a forward-tilted lever arm is the relaxed conformation
in the Po state, there is no binding penalty. Therefore, upon reaching
the capture radius $a$ around $\mb{r}_-$, the leg binds with
probability 1, and MyoV returns to the waiting state.  The fact that
backstepping in our model does not require ATP hydrolysis is
consistent with observations of ATP-independent processive backwards
stepping in the superstall regime ($F > 3$ pN)~\cite{Gebhardt06}.
For simplicity, we will not consider the superstall case
  in the present study.  In principle, our model could be generalized
  to the superstall regime by including additional kinetic pathways
  that occur under extremely large backward loads, for example power
  stroke reversal~\cite{Sellers10}.

In all of the four kinetic pathways described above, only one leg is
always bound to the actin during the diffusion step.  If the bound leg
detaches before the free leg binds, the processive run of MyoV is
terminated.  We assume a bound leg detachment rate $t_\text{d1}^{-1}$
during this process.  This completes the description of the model,
where each MyoV waiting state ends in five possible outcomes: forward
stepping, T/L foot stomps, backward stepping, or detachment of both
heads from actin.  The first four pathways bring the system back to
the waiting state, where the entire mechanochemical cycle can be
repeated, while the last ends the run.  The only parameters for which
we do not have direct experimental estimates are the strength and
direction of the power stroke constraint, $\nu_c$ and $\theta_c$, and
the binding penalty $b$.  We will be able to fit these parameters by
comparing the theoretical results to experimental data, as described
below, resulting in the values listed in Table~1.  
  Imaging studies~\cite{Walker00,Kodera10} suggest that the preferred
  Po orientation $\theta_c$ is likely to be in the vicinity of
  $60^\circ$, so this parameter could have been constrained from the
  outset.  However we have allowed it to be a free parameter since the
  angle $\theta_c$ that appears in the potential function ${\cal H}_c$
  can in principle be slightly different than the observed orientation
  of the bound leg in any particular image, which is affected by both
  thermal fluctuations and any tension that is applied to the end of
  the bound leg.  For the persistence length
$l_p$, there are estimates ranging from $l_p \approx 100$
nm~\cite{Howard96} up to 375 nm~\cite{Vilfan05}.  We use the value
$l_p = 310$ nm, based on the measurements of Moore {\em
  et. al.}~\cite{Moore04}.  From the point of view of the polymer
model, the most important characteristic of the persistence length is
that $l_p \gg L$, so the legs behave almost as rigid rods.  However,
one of the major outcomes of our theory is that precise tuning of the
parameters is not required to get efficient processive dynamics
qualitatively similar to that seen in nature.

\vspace{2em}

\tocless\subsection{Analytical theory for diffusive search times}  The
central physical quantity in our model is the first passage time to
the binding site, $t^{\pm}_\text{fp}$, which depends sensitively on
the interplay of bending stiffness ($l_p$), load force ($F$,
$\theta_F$), and power stroke constraint ($\nu_c$, $\theta_c$)
(Fig.~\ref{path}B).  The magnitude of $t^{\pm}_\text{fp}$ at a given
$F$ compared to the hydrolysis and detachment rates, along with the
size of the binding penalty, determines exactly how the system
partitions between the various kinetic pathways.

\begin{figure}
\centerline{\includegraphics[width=\columnwidth]{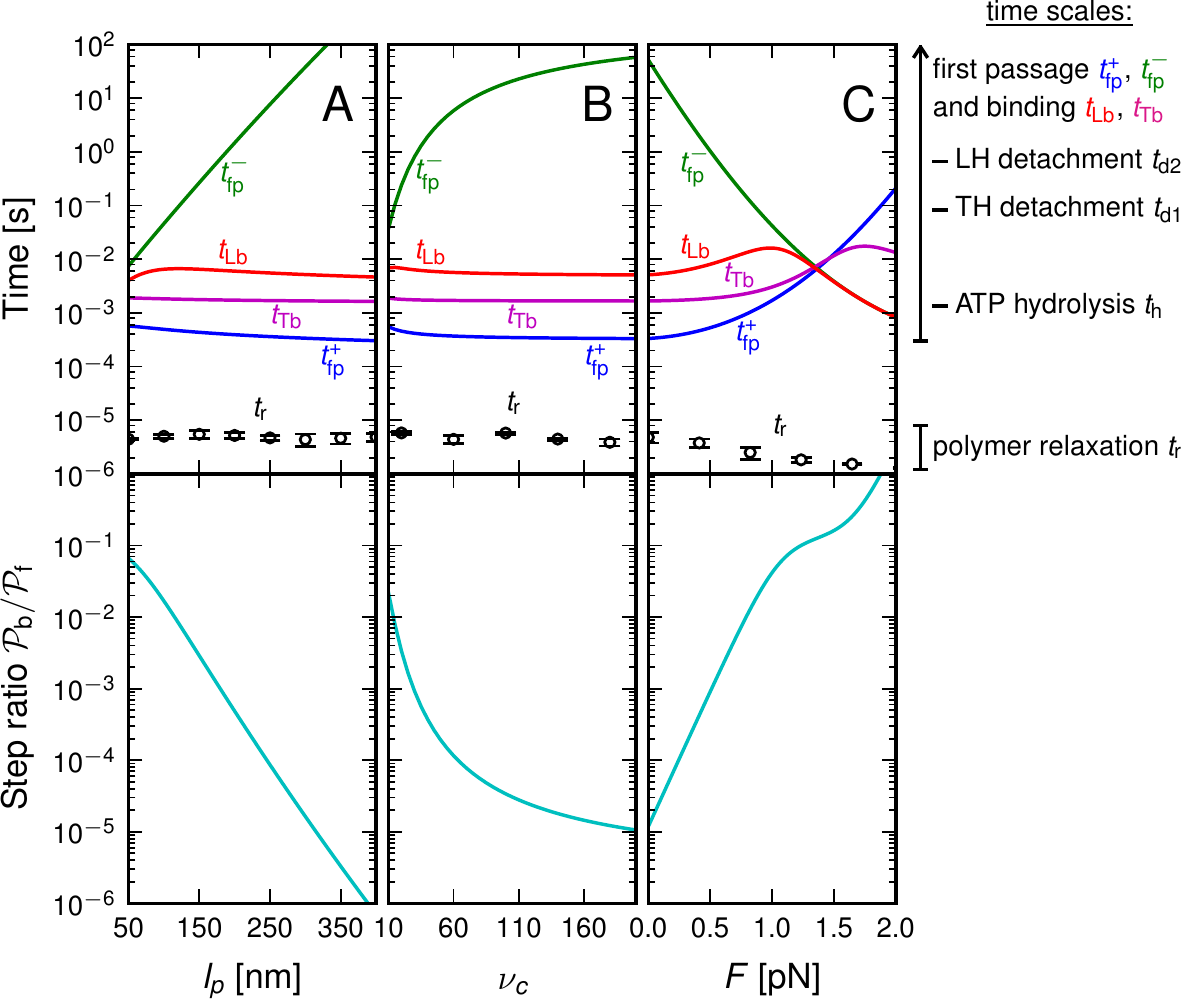}}
\caption{The top row shows theoretical predictions for the mean first
  passage times $t^\pm_\text{fp}$ to the forward (+) and backward (-)
  sites, the the mean binding times $t_\text{Lb}$, $t_\text{Tb}$ for
  the leading and trailing legs respectively, and the polymer
  relaxation time $t_\text{r}$ for the MyoV structure to equilibriate
  after the detachment of one leg.  All results except $t_\text{r}$
  are derived from the analytical theory.  The relaxation times are
  estimated using coarse-grained Brownian dynamics simulations (see SI
  for details).  On the right, the main time scales in the problem are
  summarized, with their values (or ranges) indicated for comparison.
  The bottom row shows the ratio of backward-to-forward steps, ${\cal
    P}_\text{b}/{\cal P}_\text{f}$.  For the three columns the
  quantities are plotted as one parameter is varied while all others
  are fixed at their Table~1 values: A) leg persistence
  length $l_p$; B) power stroke constraint strength $\nu_c$; C) load
  force $F$ (with $\theta_F =0$).}\label{times}
\end{figure}

Remarkably, the polymer model allows us to derive an approximate
analytical expression for $t^{\pm}_\text{fp}$ by exploiting the separation in time scales
between polymer relaxation and the diffusive search (for details see the Methods and the Supporting Information (SI)).     If $t_\text{r}$
is the relaxation time for the two-legged polymer structure to
equilibrate after one of the legs detaches, then $t_\text{r} \ll
t_\text{fp}^\pm$.  Theory and simulations show that
$t_\text{r} \approx 5$ $\mu$s for nearly rigid legs at zero load, and
becomes even smaller as $F$ increases (Fig.~\ref{times} and SI
Fig.~S1).  The value of $t_\text{r}$ is two orders of magnitude
smaller than the fastest times for first passage to the binding sites,
$t_\text{fp}^\pm \sim {\cal O}(0.1\:\text{ms})$.  Because
$t_\text{fp}^\pm/t_\text{r} \gg 1$, we can relate $t_\text{fp}^\pm$ to
the distribution ${\cal P}(\mb{r})$, the probability density of
finding the MyoV free end at position $\mb{r}$ once the system has
reached equilibrium after leg detachment,
\begin{equation}\label{e1}
t_\text{fp}^\pm \approx \frac{1}{4\pi D_\text{h} a {\cal
    P}(\mb{r}_\pm)},
\end{equation}
where $D_\text{h} = 5.7 \times 10^{-7}$ cm$^{2}$/s is the diffusion
constant of the MyoV head, estimated using the program
HYDROPRO~\cite{Ortega11} applied to the PDB structure
1W8J~\cite{Coureux04}.  
The above equation transforms the dynamical
problem of diffusive search time into one of calculating the
equilibrium end-point distribution of a tethered, two-legged
semiflexible polymer structure.  By adapting a mean field theory for
invididual semiflexible chains~\cite{Thirum98}, and noting that
contour fluctuations are small in the regime $l_p \gg L$, we obtained
an approximate but accurate analytical expression for ${\cal P}(\mb{r}_\pm)$,
taking into account both the load force on the joint and the
end-tangent constraint (Methods and SI).  Together with
Eq.~\eqref{e1}, we  have a complete description of
$t_\text{fp}^\pm$ as a function of load and the MyoV structural
parameters.  If we assume that the other events in the mechanochemical
cycle---hydrolysis and TH/LH detachment---are Poisson processes with
respective rates $t_h^{-1}$, $t_\text{d1}^{-1}$, and
$t_\text{d2}^{-1}$, the probability of each kinetic pathway can also
be derived, together with related quantities like mean run length and
velocity.  The full set of analytical equations for our model is
summarized in Table~S1 of the SI.

\vspace{2em}

\tocless\subsection{The role of diffusion in the kinetic pathway
  probabilities at zero load}  To gain an understanding of how the
structural features of MyoV influence its motility, it is
instructive to start with $F=0$.  The top
panels of Fig.~\ref{times}A,B show the first passage times
$t_\text{fp}^\pm$ as a function of $l_p$ and $\nu_c$ respectively with 
the other parameters being fixed at their Table~1
values.  Because of the power stroke constraint, there is an asymmetry
in the first passage times: $t_\text{fp}^+ < t_\text{fp}^-$ because
the center of the ${\cal P}(\mb{r})$ distribution is shifted toward
the forward binding site at $z = +\Delta$.  At $F=0$ the average
$z$-axis location of the free leg, $\mu_z = \int dr\,(\hmb{z} \cdot
\mb{r}) {\cal P}(\mb{r})$, is given by
\begin{equation}\label{e2}
\mu_z = l_p (1-e^{-\kappa}) (\coth \nu_c - \nu_c^{-1}) \cos\theta_c,
\end{equation}
where $\kappa \equiv L/l_p$, and the origin $z=0$ is at the binding
site of the attached leg.  With increasing $l_p$ and $\nu_c$, the
position $\mu_z$ increases until it saturates at the limit of a rigid
rod of length $L$ with a fixed angle $\theta_c$, $\mu_z \to L
\cos\theta_c$.  In this limit $t_\text{fp}^- \to \infty$, since it is
geometrically impossible to reach the backward binding site $z =
-\Delta$.  In the opposite limit of small $l_p$ and $\nu_c$, the
structure has greater flexibility, reaching the backward binding site
is easier, and the asymmetry is smaller.  For $\nu_c \gg 1$ and
$\kappa \ll 1$, the asymmetry parameter, $\alpha = t_\text{fp}^+/t_\text{fp}^-$
has a simple relationship to the structural
parameters,
\begin{equation}\label{e3}
\begin{split}
\alpha &= \frac{t_\text{fp}^+}{t_\text{fp}^-} \approx \exp\left(-\frac{\Delta {\cal T}}{L} \cos\theta_c + \beta \Delta F \cos\theta_F\right),\\
{\cal T} &\equiv 1 + \frac{20 \nu_c}{20+7\kappa \nu_c},
\end{split}
\end{equation}
where $\beta = 1/k_BT$ and $\Delta \approx 36$ nm is the step size.
At $F=0$ the key role in determining the degree of asymmetry is the
factor ${\cal T}$, which depends on $l_p$ and $\nu_c$ and is a
dimensionless measure of the effectiveness of the power stroke
constraint.  Larger ${\cal T}$ means a smaller $\alpha$ and greater
asymmetry.  The form of ${\cal T}$ shows that the constraint strength
$\nu_c$ by itself is insufficient to guarantee a large ${\cal T}$,
since it can be counterbalanced by a small $l_p$.  In other words, the
end-tangent constraint does not have a significant effect if the
polymer leg is too flexible.  Thus, both $\nu_c$ and $l_p$ have to be
large to create significant asymmetry.  In the
  Discussion we will highlight the relationship between ${\cal T}$ and
  important mechanical and energy scales in the system, including the
  overall compliance of the leg and the energy expended by the power
  stroke.

The asymmetry factor $\alpha$ influences kinetic pathway probabilities.
At the end of each waiting stage, there is a probability of
making a forward step (${\cal P}_\text{f}$), a backward step (${\cal
  P}_\text{b}$), an L foot stomp (${\cal P}_\text{Ls}$), and a T foot
stomp (${\cal P}_\text{Ts}$).  We plot these probabilities in
Fig.~\ref{path}D for a range of $F$.  When the time scale of leg
detachment is much larger than the binding times, the ratios of the
pathway probabilities can be expressed in terms of $\alpha$ as,
\begin{equation}\label{e4}
\frac{{\cal P}_\text{b}}{{\cal P}_\text{f}} = \frac{\alpha(1+ b
  \alpha)}{g(b+\alpha)},\qquad \frac{{\cal P}_\text{Ls}}{{\cal P}_\text{f}} = \frac{b(1+ b
  \alpha)}{g(b+\alpha)}, \qquad \frac{{\cal P}_\text{Ts}}{{\cal P}_\text{f}} = b\alpha.
\end{equation}
Note that ${\cal P}_\text{b}/{\cal P}_\text{f}$ and ${\cal
  P}_\text{Ls}/{\cal P}_\text{f}$ are inversely proportional to $g$,
the ratio of TH-to-LH detachment, which is expected since backward
steps and L foot stomps can only occur when the LH detaches.  The
binding penalty $b$ enters into all the ratios because it influences
the likelihood of T/L foot stomping, which compete with the
backward/forward stepping pathways.  The bottom panels of
Fig.~\ref{times}A,B show the variation of ${\cal P}_\text{b}/{\cal
  P}_\text{f}$ at $F=0$ as $l_p$ and $\nu_c$ are varied.  We find that
${\cal P}_\text{b}/{\cal P}_\text{f}$ decreases as either variable
increases, due to larger ${\cal T}$ in Eq.~\eqref{e3} resulting in
smaller $\alpha$.  Experimentally, MyoV exhibits negligible backstepping
at zero load, ${\cal P}_\text{b}/{\cal P}_\text{f} \lesssim 1
\%$~\cite{Kad08}.  In order to achieve this extreme unidirectionality,
${\cal T}$ (or equivalently both $\nu_c$ and $l_p$) should be
sufficiently large, an issue we will return to in the Discussion
section when we examine the global constraints on the structural features
of the motor.  Along with backsteps, T foot stomps are also negligible
at $F=0$ for small $\alpha$, since ${\cal P}_\text{Ts}/{\cal
  P}_\text{f}\propto \alpha$.  As $\alpha \to 0$, the only ratio that has
a nonzero limit is ${\cal P}_\text{Ls}/{\cal P}_\text{f} \to g^{-1}$.
Qualitatively similar behavior was observed in the high speed AFM
experiments~\cite{Kodera10}, where the TH rarely detached without
resulting in a forward step. On the other hand, essentially every time
the LH detaches, it will rebind to its original location (L stomp)
since the power stroke constraint prevents it from reaching the
backward site.  For example, in the $F=0$ slice of Fig.~\ref{path}D,
${\cal P}_\text{f} \approx g/(1+g) = 0.89$ and ${\cal P}_\text{Ls}
\approx 1/(1+g) = 0.11$. The other pathways not contribute
significantly.

\vspace{2em}

\tocless\subsection{Binding dynamics and the average step trajectory at zero
  load} The mean times $t_\text{Tb}$ and $t_\text{Lb}$ for the TH and
LH to bind after detachment (irrespective of the binding site) are
related to $t_\text{fp}^\pm$ as
\begin{equation}\label{e4b}
t_\text{Tb} = t_\text{h} + \frac{t^+_\text{fp}}{1+b\alpha},\qquad t_\text{Lb} = \frac{t^+_\text{fp}}{b+\alpha}.
\end{equation}
These binding times are plotted in the top panels of
Fig.~\ref{times}A,B as a function of $l_p$ and $\nu_c$.  The detached
TH has to undergo hydrolysis before rebinding, so $t_\text{Tb} >
t_\text{h}$.  For the parameters in Table~1, $t^+_\text{fp}
= 0.3$ ms and $t_\text{h} = 1.3$ ms at $F=0$, so hydrolysis is the
rate-limiting step for TH binding.  As noted above, T foot
stomping is infrequent in this case, so the binding events
contributing to $t_\text{Tb}$ are almost exclusively forward steps.
We note, {\it en passant}, that our value for $t^+_\text{fp}$ agrees well the $F=0$
result of Brownian dynamics simulations~\cite{Craig09}, further 
validating the analytical model for the diffusive search.

\begin{figure}
\centerline{\includegraphics[width=\columnwidth]{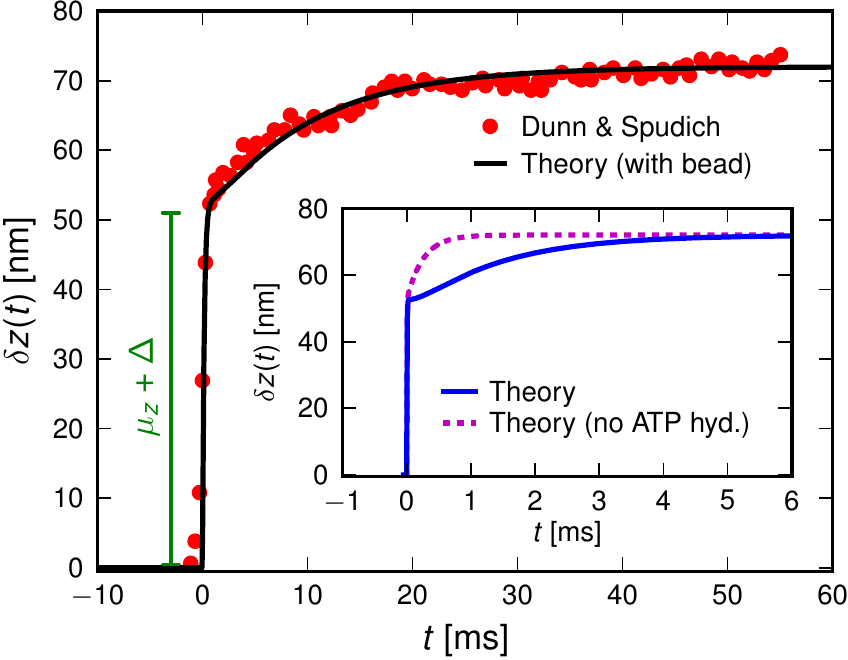}}
\caption{The mean step trajectory $\langle \delta z(t) \rangle$ of the
  detached head along the actin filament at zero load.  Red dots are
  the experimental results of Dunn \& Spudich~\cite{Dunn07}, obtained
  by tracking a gold nanoparticle attached near the end of the MyoV
  lever arm.  A fast rise occurs over a distance $\mu_z + \Delta$,
  resulting from the polymer structure relaxing to equilibrium after
  TH detachment.  The more gradual rise that follows
  corresponds to the diffusive search for the forward binding site.
  The solid curve is the theoretical prediction, corrected for the
  slowing down of relaxation and first passage dynamics due to the
  particle.  Inset: the result of the original theory without
  the correction (solid curve), compared with a variant of the theory
  where ATP hydrolysis is removed as a condition for the TH
  to bind (dashed curve).}\label{avgstep}
\end{figure}

A closely related quantity to the mean binding time is the cumulative
probability that the head has bound to a particular binding site at
time $t$ after detachment.  For the TH the probability ${\cal
  P}^{\pm}_\text{Tb}(t)$ for the site $\mb{r}_\pm$ is given by:
\begin{equation}\label{e5}
\begin{split}
{\cal P}_\text{Tb}^+(t) &= \frac{t_\text{h}\left(1-e^{-t/t_\text{h}}\right) - t_\text{fp}^+(1+b\alpha)^{-1} \left(1-e^{-t/t_\text{fp}^+}\right)}{t_\text{h}(1+b\alpha)-t_\text{fp}^+},\\
{\cal P}_\text{Tb}^-(t) &= b \alpha {\cal P}_\text{Tb}^+.
\end{split}
\end{equation}
From ${\cal P}^{\pm}_\text{Tb}(t)$ we can calculate an experimentally
measurable quantity, the average distance traveled by the free end
along the $z$-axis after detachment, $\langle \delta z(t) \rangle =
\langle z(t) -z(0) \rangle$, where $z(0) = -\Delta$.  The result is:
\begin{equation}\label{e6}
\begin{split}
\langle \delta z(t) \rangle &= (\mu_z+\Delta)(1-{\cal
  P}^+_\text{Tb}(t) - {\cal P}^-_\text{Tb}(t))(1-e^{-t/t_\text{r}})\\
&\qquad + 2\Delta {\cal
  P}^+_\text{Tb}(t).
\end{split}
\end{equation}
The first term represents the contribution from the ensemble of
trajectories where the TH is still unbound: a fast polymer relaxation
over time $t_\text{r}$ from the initial point at $z(0) = -\Delta$ to
the equilibrium average position $\mu_z$ [Eq.~\eqref{e2}].  The second
term represents the fraction of the ensemble where the TH has
successfully bound to the forward site, which eventually corresponds
to the entire ensemble for sufficiently large $t$.  Thus, $\delta
z(t)$ has two regimes, as shown in Fig.~\ref{avgstep}: a steep rise to
$\mu_z + \Delta$ on timescales $t \lesssim t_\text{r}$, followed by a
slower ascent to the full step distance $2 \Delta$.  Dunn and Spudich
have measured $\langle \delta z(t) \rangle$ for MyoV by attaching a
40-nm-diameter gold nanoparticle near the end of one lever
arm~\cite{Dunn07}.  Observing the particle through dark-field imaging,
they aligned and averaged 231 individual step trajectories to produce
the $\langle \delta z(t) \rangle$ data points shown in
Fig.~\ref{avgstep}.  Because the nanoparticle is sufficiently large
that its hydrodynamic drag will slow down the relaxation and diffusive
dynamics, we included a time rescaling factor $B$ into the theory to
account for the effect of the bead: $t_\text{fp}^+ \to B
t_\text{fp}^+$, $t_\text{r} \to B t_\text{r}$.  The theory agrees well
with experiment for $B = 29$ and $\theta_c = 60^\circ$.  The fitted
value of $\theta_c$ is based on setting the experimentally measured
steep rise, $\approx 52$ nm, equal to $\mu_z + \Delta$, with $\mu_z$
given by Eq.~\eqref{e2}.  The $\theta_c$ value is insensitive to the
precise value of $\nu_c$ or $l_p$ (assuming we are in the $\nu_c \gg
1$ and $\kappa \ll 1$ regime), as well as the time rescaling $B$.  In
the experiment the relaxation time for the steep rise was faster than
the equipment time resolution of 320 ms.  In our theory the rescaled
relaxation time $B t_\text{r} \approx 145$ ms, which satisfies this
upper bound.  After the steep rise, the remaining $\approx 20$ nm
ascent to the full step distance is determined by the diffusive search
and binding to the forward site.  According to
Eqs.~\eqref{e5}-\eqref{e6}, this part of the step involves two time
scales, $t_\text{h}$ and $t^+_\text{fp}$.  Though
  $t^+_\text{fp} \approx 0.33$ ms is smaller than $t_\text{h} = 1.3$
  ms, the rescaled $B t^+_\text{fp} = 9.7\:\text{ms} > t_\text{h}$, so
  in this particular case hydrolysis is not rate limiting.

  However, by changing the ATPase properties of the motor head, one
  can experimentally observe the role of hydrolysis in the binding
  kinetics.  The nanoparticle tracking results described above are for
  MyoV with essential light chain LC1sa at the lever arm binding site
  closest to the motor head, and calmodulin along the remainder of the
  arm.  We will denote this type as MyoVelc.  Dunn and Spudich also
  studied a variant with only calmodulin (MyoVcam) that has very
  different ATPase rates.  As shown in an earlier bulk
  study~\cite{delaCruz00}, for MyoVelc the reverse hydrolysis
  rate $t_{-\text{h}}^{-1}$ is negligible compared to the forward rate
  ($t_\text{h}^{-1} = 750$ s$^{-1}$ from Table 1) with
  $t_\text{h}/t_{-\text{h}} < 0.1$.  In contrast, for MyoVcam the
  forward rate is more than four times slower, $t_\text{h}^{-1} = 162$
  s$^{-1}$, and the reverse rate is substantial, $t_{-\text{h}}^{-1} =
  216$ s$^{-1}$~\cite{delaCruz00}.  With non-negligible
  $t_{-\text{h}}^{-1}$ and the bead rescaling factor $B$,
  Eq.~\eqref{e4b} for the TH binding time becomes
\begin{equation}\label{e6b}
  t_\text{Tb} = t_\text{h} + \frac{B t^+_\text{fp}}{1+b\alpha}\left(1+\frac{t_\text{h}}{t_{-\text{h}}}\right).
\end{equation}
By substituting the $t_\text{h}^{-1}$ and $t_{-\text{h}}^{-1}$
estimates for MyoVcam from Ref.~\cite{delaCruz00}, while keeping all
other parameters the same, Eq.~\eqref{e6b} predicts a MyoVcam binding
rate $t_\text{Tb}^{-1} = 35$ s$^{-1}$, about 2.6 times slower than for
MyoVelc, where $t_\text{Tb}^{-1} = 91$ s$^{-1}$.  Dunn and Spudich
estimated the rebinding rates from the nanoparticle trajectories, and
found a similar threefold decrease between the MyoVelc and MyoVcam
systems, from $180 \pm 50$ s$^{-1}$ down to $60 \pm 15$
s$^{-1}$~\cite{Dunn07}.  The experimental rebinding rates are faster
than the theoretical ones, which may be due in part to the fact that
experimentally rebinding is not directly observed, but only
approximately inferred from where the $\delta z(t)$ trajectory covers
the full distance $2\Delta$ to the forward site.  The myosin head
could still diffuse near $2\Delta$ for some time without binding, and
this could be indistiguishable from a binding event due to the
intrinsic noise in the trajectory.  However the general slowdown seen
in the experiment is reproduced in the theory, and highlights the
interplay of hydrolysis and diffusion times in the binding dynamics.

The hydrolysis rate would also play a greater role if the impediment
of the attached bead were removed.  For the MyoVelc case, with a bead
factor $B=29$, the free end has enough time to hydrolyze before
finding the forward binding site.  Thus, the decay after the steep
rise is mainly single exponential in Fig.~\ref{avgstep}, with a
characteristic time $B t^+_\text{fp}$.  If a future experiment were to
measure $\langle \delta z(t)\rangle$ without slowing down the
diffusion, we should see the average step shape shown in the inset of
Fig.~\ref{avgstep}, predicted by the theory for $B=1$.  There is a more
gradual, double-exponential, decay after the steep rise, reflecting
both the $t_\text{h}$ and $t_\text{fp}^+$ time scales.  For
comparison, we also show the results of the theory without ATP
hydrolysis as a precondition for binding, in order to emphasize the
change in the $\langle \delta z(t)\rangle$ shape due to $t_\text{h}$.

\begin{figure}
\centerline{\includegraphics[width=\columnwidth]{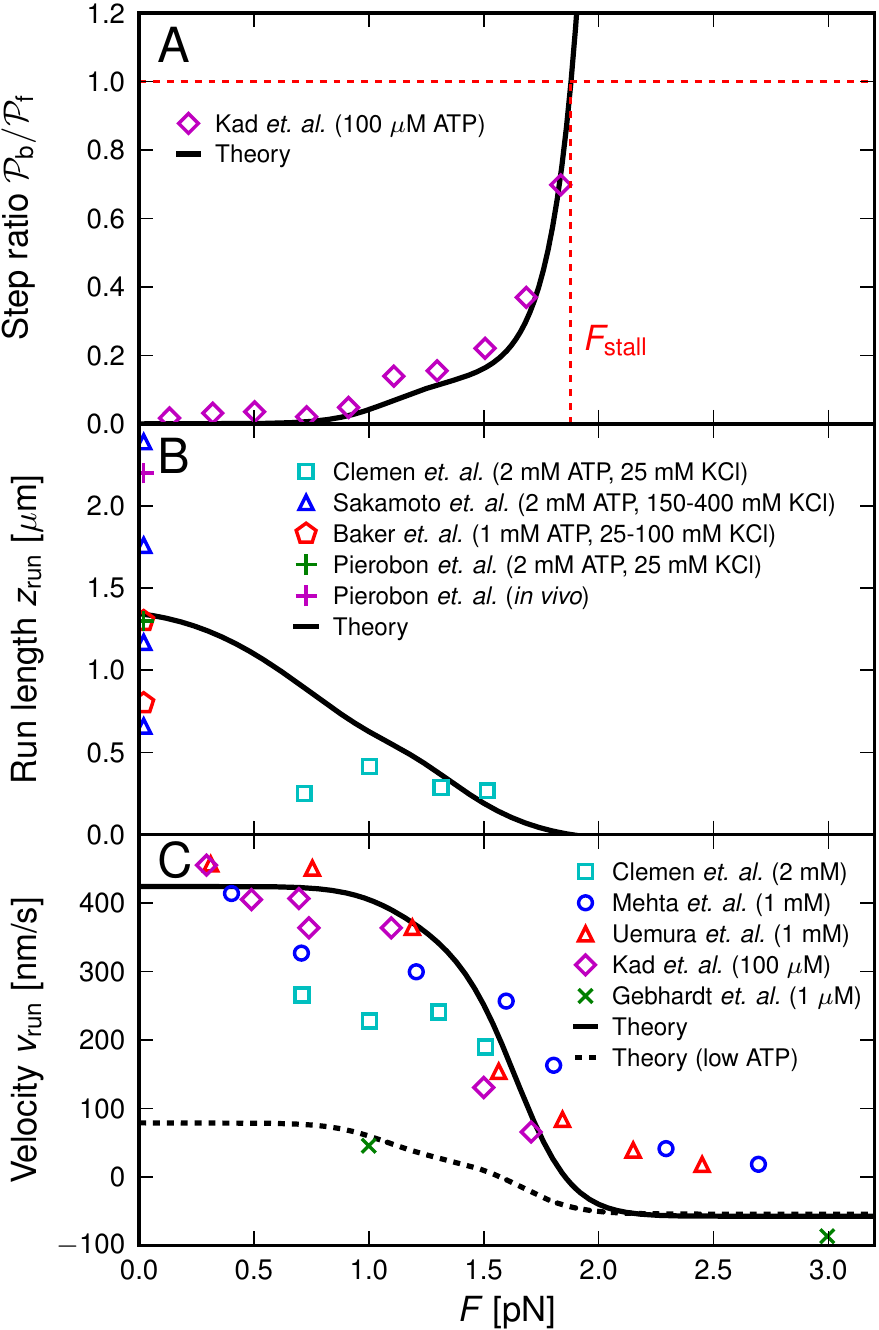}}
\caption{Comparison of the theory predictions (solid curves, with
  parameters in Table~1) to experimental results (symbols)
  as a function of load force $F$ (with $\theta_F =0$).  For the
  legends, the first and second terms in the parentheses correspond to
  experimental ATP and KCl concentrations respectively.  Where the KCl
  concentration is not indicated, the value is 25 mM.  A) The ratio of
  backward-to-forward steps, ${\cal P}_\text{b}/{\cal P}_\text{f}$,
  compared to the data of Ref.~\cite{Kad08}.  B) Run length
  $z_\text{run}$, compared to the data of
  Refs.~\cite{Clemen05,Sakamoto00,Baker04,Pierobon09}. C) Velocity
  $v_\text{run}$, compared to the data of
  Refs.~\cite{Clemen05,Mehta99,Uemura04,Kad08,Gebhardt06}.  The dashed
  curve corresponds to a modified version of the theory which accounts
  for the low ATP concentration in the Gebhardt~{\em et. al.}
  experiment~\cite{Gebhardt06} (see main text).}\label{exp}
\end{figure}

Experimentally, one can also study the average $z$-axis trajectory of
the center-of-mass, for example in a single-molecule bead
assay~\cite{Cappello07}.  The results are essentially similar, but
with the distances above are halved: we have a fast $\approx 26$ nm
rise corresponding to the power stroke, detachment, and polymer
relaxation, and the remaining slow $\approx 10$ nm ascent due to
diffusive search and binding, giving a combined $36$ nm center-of-mass
step.

\vspace{2em}

\tocless\subsection{Run length at zero load}  The final observable quantity
of interest at $F=0$ is the mean run length along the actin filament.
Assuming $t_\text{d1} \gg t_\text{Lb},\: t_\text{Tb}$, the average run
length $z_\text{run}$ at any $F$ is given by:
\begin{equation}\label{e7}
\begin{split}
z_\text{run} &= v_\text{run} t_\text{run}, \qquad v_\text{run} \approx \frac{\Delta}{t_\text{d1}}\left( \frac{1}{1+b\alpha} - \frac{\alpha}{g(b+\alpha)}\right),\\
t_\text{run} &\approx \frac{g t_\text{d1}^2}{t_\text{Lb}+ g t_\text{Tb}},
\end{split}
\end{equation}
where $v_\text{run}$ and $t_\text{run}$ are the mean run velocity and
duration.  The positive and negative terms in $v_\text{run}$ are
contributions from forward and backward stepping respectively.
Experimental estimates for $z_\text{run}$ at $F=0$, plotted on the
left edge of Fig.~\ref{exp}B, vary over a wide range from $0.7-2.4$
$\mu$m~\cite{Sakamoto00,Baker04,Pierobon09}, most likely due to
different measurement conditions (particularly the KCl concentration
of the buffer).  We choose as a representative value $z_\text{run} =
1.3$ $\mu$m, which allows us to use Eq.~\eqref{e7} at $F=0$ to solve
for the binding penalty parameter, $b = 0.065$.  This can be done
since $\alpha \ll 1$ at zero load, and substituting $\alpha =0$ in
Eq.~\eqref{e7} leads to an expression that is roughly independent of
$\nu_c$ for large $\nu_c$. Thus we have fit two of the free
parameters, $\theta_c$ and $b$, by comparison with experimental values
for the rise $\mu_z + \Delta$ and the run length $z_\text{run}$
respectively.  The final free parameter, $\nu_c$, will be fit by
comparison to the stall force, discussed in the next section.

\vspace{2em}

\tocless\subsection{Load dependence of the kinetic pathways and a simple
  formula for stall force}  When a backwards force is applied to
MyoV, it counteracts the bias due to the power stroke constraint,
bending the bound leg and shifting the equilibrium away from the
forward binding site.  We see this directly in Eq.~\eqref{e3} for
$\alpha$, where the $\beta \Delta F \cos\theta_F$ term in the
exponential has the opposite sign of the $-\Delta {\cal T} L^{-1}\cos
\theta_c$ contribution from the constraint.  Thus, $\alpha$ increases
rapidly with increasing $F$, eventually becoming greater than 1,
meaning that reaching the backward site is faster than reaching the
forward one.  Fig.~\ref{times}C plots $t_\text{fp}^\pm$ and the leg
binding times as a function of $F$ for the parameter set in
Table~1.  The changeover from $\alpha <1$ to $\alpha >1$ occurs
near $F=1.4$ pN.  The corresponding pathway probabilities are in
Fig.~\ref{path}D.  With increasing force each leg changes its primary
kinetic pathway.  TH detachment, which almost always leads to forward
stepping at small $F$, instead leads to T foot stomping at high $F$.
Similarly LH detachment resulted in mainly L foot stomps at low $F$,
but leads to backward stepping at high $F$.  Thus, application of a
resistive load totally alters the partioning between the kinetic
pathways.

At the stall force, $F_\text{stall}$,  the probabilities of
backward and forward stepping are equal, and the mean MyoV velocity
goes to zero.  Setting ${\cal P}_\text{b}/{\cal P}_\text{f}$ from
Eq.~\eqref{e4} equal to 1, substituting $\alpha$ from Eq.~\eqref{e3},
we obtain:
\begin{equation}\label{e8}
\begin{split}
F_\text{stall} &= \frac{{\cal T} \cos\theta_c}{\beta L \cos\theta_F}\\
&\quad + \frac{1}{\beta\Delta \cos\theta_F}\log \frac{g-1+\sqrt{(g-1)^2 +4gb^2}}{2 b}\\
&\equiv F_\text{stall}^\text{p} + F_\text{stall}^\text{c},
\end{split}
\end{equation}
where the power stroke effectiveness ${\cal T}$ is defined in
Eq.~\eqref{e3} in terms of $l_p$, $L$, and $\nu_c$.  The stall force
has two main contributions. (i) The first term
$F_\text{stall}^\text{p}$ is due to the power stroke constraint,
depending on ${\cal T}$ and $\theta_c$, and thus the structural
parameters which determine ${\cal T}$.  Larger ${\cal T}$ and smaller
$\theta_c$ both act to shift the free end probability distribution
closer to the forward site, impeding backstepping and contributing to
a larger $F_\text{stall}$.  (ii) The second term
$F_\text{stall}^\text{c}$ arises from two properties of MyoV head
chemistry: the gating ratio $g$ which controls how often the trailing
head detaches relative to the leading head, and the binding penalty
due to incorrect head orientation near the binding site.  Increasing
$g$ makes detachment of the LH less common.  Since backstepping
requires LH detachment it will also become less probable.  The
importance of $b$ is related to the Pr orientation penalty, which
makes binding to the backward site less favorable.  Larger $g$ or
smaller $b$ reduces ${\cal P}_\text{b}/{\cal P}_\text{f}$ at any given
$F$, thus increasing $F_\text{stall}$.  If there were no gating
asymmetry (the ratio $g=1$) then the contribution
$F_\text{stall}^\text{c}$ vanishes.  

The optical trap experiment of Kad {\em et. al.}~\cite{Kad08} yielded
${\cal P}_\text{b}/{\cal P}_\text{f}$ as a function of $F$.  The data
are plotted in Fig.~\ref{exp}A, corresponding to an estimated
$F_\text{stall} \approx 1.9$ pN. Using this experimental value of
$F_\text{stall}$, and assuming for simplicity $\theta_F = 0$ or a pure
backwards load, we get $\nu_c = 184$ by solving Eq.~\eqref{e8}.
In the Discussion we will return to the magnitude of
  $\nu_c$ in the broader context of stiffness and energetics within
  the myosin motor family.  The theoretical curve in Fig.~\ref{exp}A
is in good agreement with the experimental datapoints over the entire
measured $F$ range.  Backstepping is mostly suppressed for $F \lesssim
1$ pN, and then rapidly increases until the stall point.

\vspace{2em}

\tocless\subsection{Run length and velocity under load} The change in
kinetic pathways with $F$ manifests itself in two other observables,
the mean run length $z_\text{run}$ and velocity $v_\text{run}$, which
both decrease to zero as the stall force is approached.
In Fig.~\ref{exp}B and C we show various experimental results for these two
quantities as a function of $F$, together with the theoretical
prediction [Eq.~\eqref{e7}].  Aside from one exception mentioned
below, all the experiments were done at saturating ATP ($\gtrsim 100$
$\mu$M).  Despite the scatter in the experimental values, the theory
reproduces the overall trends well.  The motor functions nears its
unloaded ($F=0$) velocity of $v_\text{run}=414$ nm/s ($\approx \Delta t^{-1}_{d1}$) for
small forces, and then slows down noticeably for $F \gtrsim
1$ pN, as the proportion of backsteps increases.  The extrapolated
force at which the velocity goes to zero is another way to
estimate the stall force, and the experiments show MyoV stalling
in the range of $F \approx 1.9-3$ pN.

Above the stall force, the theory predicts a small net negative
velocity, since back steps outnumber the forward steps.  Although the
present theory will likely require modifications at very high forces
far into the superstall regime, we can tentatively compare our results
to those of Gebhardt {\em et. al.}~\cite{Gebhardt06} at $F=1$ pN and
$F=3$ pN (green crosses in Fig.~\ref{exp}C), where the latter data
point was just above stall, and exhibited a small negative velocity
$\approx -90$ nm/s.  In this case the ATP concentration is 1 $\mu$M,
which makes ATP binding the rate limiting step in TH detachment.  To
accommodate this, we set $t_{d1}^{-1} = 2.2$ s$^{-1}$, which is the
binding rate at 1 $\mu$M ATP estimated from the experimental
kinetics~\cite{Gebhardt06}.  With this single modification, the theory
gives the dashed curve in Fig.~\ref{exp}C, which roughly captures the
velocities both below and above stall.  Taken together, the comparison
between the theory and a number of experimental results shows that our
predictions agree with measurements remarkably well.

\vspace{2em}

\tocless\section{Discussion}

\tocless\subsection{Constraints on MyoV structural and binding parameters}
\begin{figure}
\centerline{\includegraphics[width=\columnwidth]{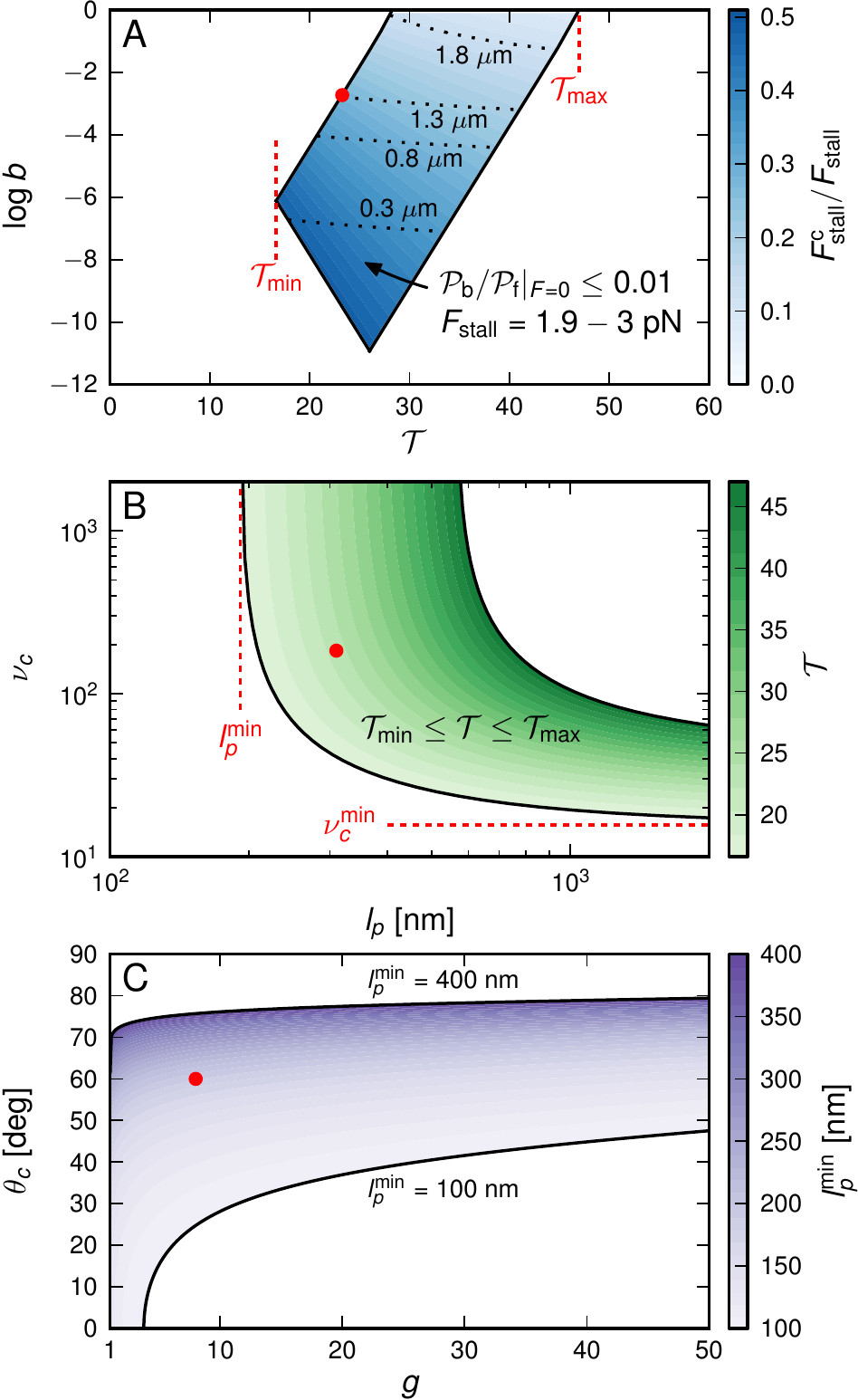}}
\caption{Exploring the design space for MyoV satisfying the
  constraints that ${\cal P}_\text{b}/{\cal P}_\text{f}<0.01$, and
  that the stall force $F_\text{stall}$ be in the range $1.9-3$ pN.
  The red dot in each panel corresponds to the parameter set in
  Table~1.  A) The blue shaded region shows the allowed
  values for the binding penalty $b$ and power stroke effectiveness
  ${\cal T}$ [Eq.~\eqref{e3}].  The intensity of the shading indicates
  the fraction $F^\text{c}_\text{stall}/F_\text{stall}$, where
  $F^\text{c}_\text{stall}$ is the contribution of head chemistry to
  the total stall force [Eq.~\eqref{e8}].  The labeled black dotted
  lines correspond to loci of constant run length $z_\text{run}$.  The
  blue shaded region falls entirely within the range ${\cal
    T}_\text{min} =16.6$ to ${\cal T}_\text{max}=47.0$ along the
  ${\cal T}$ axis.  B) The green shaded region corresponds to those
  values of persistence length $l_p$ and power stroke strength $\nu_c$
  which yield ${\cal T}$ in the range ${\cal T}_\text{min}$ to ${\cal
    T}_\text{max}$.  The intensity of shading indicates the magnitude
  of ${\cal T}$.  Below the value $l_p^\text{min} = 192$ nm and
  $\nu_c^\text{min} = 15.6$ it is impossible to satisfy the bounds on
  ${\cal T}$. C) The purple shaded region corresponds to values of the
  gating ratio $g$ and power stroke constraint angle $\theta_c$ that
  yield $l_p^\text{min}$ between $100-400$ nm, with the shading
  intensity proportional to $l_p^\text{min}$.}\label{const}
\end{figure}

MyoV walks nearly unidirectionally at zero load, and can persist
against backward loads up to the stall force.  Is the system robust to
variations in the parameter space?  To make the question concrete, we
can ask under what conditions does MyoV fulfill two requirements
for processive motion and the ability to sustain load: (i) the
backward-to-forward step ratio at zero load, ${\cal P}_\text{b}/{\cal
  P}_\text{f} \le \epsilon$; (ii) the stall force $F_\text{stall}$
falls in some range $F^\text{min}_\text{stall}$ to
$F^\text{max}_\text{stall}$ when the resistive load is applied
parallel to the actin axis ($\theta_F = 0$ in Fig.~\ref{path}B).  We
choose experimentally motivated values of $\epsilon = 0.01$,
$F_\text{stall}^\text{min} = 1.9$ pN~\cite{Kad08},
$F_\text{stall}^\text{max} = 3.0$ pN~\cite{Mehta99,Veigel02,Uemura04}.
From Eqs.~(\ref{e3}, \ref{e4}, \ref{e8}), these two conditions
are satisfied within the blue shaded area of Fig.~\ref{const}A, which
plots a $\log b$ vs. ${\cal T}$ slice of the parameter space, with
fixed $\theta_c$, $\Delta$, and $g$.  Along the ${\cal T}$ axis, the
region has minimal and maximal boundaries,
\begin{equation}\label{e10}
\begin{split}
{\cal T}_\text{min} &= \frac{L}{2 \Delta
  \cos\theta_c}\Biggl(\beta\Delta F^\text{min}_\text{stall}\\
&\quad +
\log \frac{(1-\epsilon g)e^{\beta \Delta F^\text{min}_\text{stall}}-1+g}{g (\epsilon (g-1)e^{\beta \Delta F^\text{min}_\text{stall}} +1 - \epsilon g)} \Biggr)\\
& = 16.6,\\
{\cal T}_\text{max} &= \frac{L}{\Delta
  \cos\theta_c}\left(\beta\Delta F^\text{max}_\text{stall} -
\log g\right) = 47.0,
\end{split}
\end{equation}
where the numerical values are computed for the specific parameters in
Table~1.  If ${\cal T} < {\cal T}_\text{min}$ or ${\cal T} >
{\cal T}_\text{max}$ there is no value of $b$ where conditions (i) and (ii) are
satisfied simultaneously.  A density plot of ${\cal T}$ in terms of
$\nu_c$ and $l_p$ is shown in Fig.~\ref{const}B, with the ${\cal
  T}_\text{min} \le {\cal T} \le {\cal T}_\text{max}$ region shaded in
green.  Asymptotically, this region is bounded by a minimum
persistence length $l_p^\text{min}$ for $\nu_c \to \infty$, and a
minimum constraint strength $\nu_c^\text{min}$ for $l_p \to \infty$:
\begin{equation}\label{e11}
\begin{split}
l_p^\text{min} &= \frac{7L}{20}({\cal T}_\text{min}-1) = 192\:\text{nm},\\
 \nu_c^\text{min} &= {\cal T}_\text{min} -1 = 15.6.
\end{split}
\end{equation}
Having $l_p$ and $\nu_c$ above these two minima constitute necessary,
but not sufficient conditions, for ${\cal T}$ to fall between ${\cal
  T}_\text{min}$ and ${\cal T}_\text{max}$.  Physically, ${\cal T}$
represents the effectiveness of the power stroke constraint, which is
directly related to $l_p$ and $\nu_c$ through Eq.~\eqref{e3}.  We thus
see that motor function with the given specifications requires a
certain minimal power stroke effectiveness, which cannot be achieved
unless both the persistence length of the lever arms and the strength
of the end-tangent constraint are both large enough.  If either $l_p$
or $\nu_c$ is too small, backstepping becomes more frequent at zero
load, and it is easier to bend the bound leg backward, resulting in
stall being reached at smaller force magnitudes.

The bounds on ${\cal T}$ in Eq.~\eqref{e10} also depend on the gating
ratio $g$ and Po orientation $\theta_c$ (Fig.~\ref{path}B), which we
illustrate in Fig.~\ref{const}C by plotting the density of
$l_p^\text{min}$ (related to ${\cal T}_\text{min}$ through
Eq.~\eqref{e11}) in terms of $g$ and $\theta_c$.  By showing only the
range $l_p^\text{min} = 100-400$ nm, comparable with estimates of the
lever arm persistence length~\cite{Howard96,Vilfan05}, we see there
are constraints on the angle $\theta_c$ that vary depending on $g$.
Angles too close to 90$^\circ$ give insufficient forward bias, and
have to be compensated for by an unrealistically stiff lever arm
$l_p^\text{min} > 400$ nm.  As $\theta_c$ decreases, $l_p^\text{min}$
decreases, since the stronger forward bias means that one can use
progressively more flexible lever arms and still get efficient
motility and resistance to load.

The parameter range where the motility conditions (i) and (ii) are
simultaneously satisfied (the shaded region in Fig.~\ref{const}A), is
broad, encompassing a wide swathe of possible $b$ values.  To restrict
the parameters further, we can specify that MyoV exhibit a certain run
length.  The dotted lines in Fig.~\ref{const}A are loci of
constant $z_\text{run}$, with the star marking the parameter set in
Table 1 (where $z_\text{run} = 1.3$ $\mu$m and $F_\text{stall} = 1.9$
pN).  Even with this restriction, we still have a range of possible
${\cal T}$ values at each $z_\text{run}$, which corresponds to a
region in the space of $l_p$ and $\nu_c$.  Interestingly, the system
has a degree of robustness against changes in the structural
parameters, and can meet the basic requirements for function with high
duty ratio assuming $l_p$ and $\nu_c$ yield a ${\cal T}$ within the
allowed range.

\vspace{2em}

\tocless\subsection{Relative contributions of power stroke and head chemistry
  to the stall force magnitude}  Though the emphasis in the preceding
section has been on the structural parameters, it is important to note
the complementary role of head chemistry (determined by the nucleotide
state of MyoV) in producing the observed stall force.  If trailing and
leading head detachment were equally probable ($g=1$), the
$F_\text{stall}^\text{c}$ term in Eq.~\eqref{e8} would be zero, and
$F_\text{stall} = F_\text{stall}^\text{p}$.  From the definition of
$\alpha$ in Eq.~\eqref{e3}, one can see that $F_\text{stall}^\text{p}$
is the force magnitude at which $\alpha=1$.  In other words, at $g=1$
the only condition for stall is that the first passage times to the
forward and backward sites are equal.  In fact, the value of
$F_\text{stall}^\text{p}$ arises from a simple force balance: when the
component $F \cos\theta_F$ of the backward load along the $z$-axis
equals $({\cal T}/\beta L) \cos\theta_c$, the $z$ component of an
effective forward force $({\cal T}/\beta L)$ oriented along the power
stroke constraint direction.  This is another way of interpreting the
power stroke parameter ${\cal T}$, relating it to a counteracting
force on the joint to oppose the load.  When the two forces are equal,
there is no bias either forward or backward, and $\alpha=1$.

Head chemistry changes the picture, by making leading head detachment
less frequent ($g>1$) and introducing a binding penalty ($b<1$) for
the wrong head orientation at the binding site.  A small $b$ parameter
reduces the probability of T foot stomping, which otherwise would compete
more easily with forward stepping at large loads and reduce the
likelihood of the latter.  This is the beneficial role of the recovery
stroke highlighted in Ref.~\cite{Shiroguchi11}.  The outcome is an
additional contribution $F_\text{stall}^\text{c}$ to $F_\text{stall}$,
which means stall is delayed until we reach a value of $\alpha > 1$.  In
order for backstepping to be as likely as forward stepping, it is not
enough to make the first passage times to the two binding sites equal.
We have to make $t_\text{fp}^-$ fast enough compared to
$t_\text{fp}^+$ to compensate for the gating and binding biases.  To
illustrate the significance of the $F_\text{stall}^\text{c}$
contribution over the allowed parameter range, we use the intensity of
the shading in Fig.~\ref{const}A to represent
$F^\text{c}_\text{stall}/F_\text{stall}$, the fraction of the stall
force magnitude due to the head chemistry term.  The fraction values
vary from $\approx 0.08 - 0.49$, with the parameter set in
Table~1 giving $F^\text{c}_\text{stall}/F_\text{stall} =
0.28$.  Though the power stroke term always dominates, head chemistry
has a smaller but non-negligible role in helping MyoV move forward
under load.

\vspace{2em}

\tocless\subsection{Relation of the power stroke constraint strength to
    myosin stiffness and thermodynamic efficiency}

  The mechanical compliance of MyoV under load is determined both by the
  bending stiffness of the lever arm $l_p$, and the strength of the
  effective end-tangent constraint $\nu_c$.  The latter arises at a
  molecular level from the bending stiffness of the flexible joint
  between the motor head and lever arm domains.  If we suppose this
  joint involves subdomains (i.e. the converter region of the motor
  head) on length scales $\sim 1$ nm, then $\nu_c = 184$ corresponds
  to a persistence length of $\sim 184$ nm for the head-arm joint,
  which is reasonable, since it is the same order of magnitude as the
  persistence length $l_p = 310$ nm of the lever arm itself.  

  The complex coupling between these two different bending rigidities
  is reflected in the power stroke effectiveness parameter ${\cal T}$,
  which depends nonlinearly on both $l_p$ and $\nu_c$.  In fact, one
  can approximately relate ${\cal T}$ to the overall compliance of the
  head-arm system.  For large $l_p$, where the arms are nearly rigid
  rods, the backward force ($\theta_F = 0$) required to keep the end
  of the bound leg at an angle $\theta_c^\prime > \theta_c$ (see
  Methods) is
\begin{equation}\label{e12}
F \approx \frac{{\cal T}}{\beta L} \frac{\sin(\theta_c^\prime-\theta_c)}{\sin\theta_c^\prime}.
\end{equation}
The horizontal $\delta z$ displacement
corresponding to the angular displacement between $\theta_c^\prime$
and $\theta_c$ is $\delta z \approx L(\cos\theta_c -
\cos\theta_c^\prime)$.  For $\theta_c^\prime = 60^\circ - 120^\circ$,
the rough angular range during the motor cycle, $F$ scales almost
linearly with $\delta z$, with a slope $k \approx {\cal T}/\beta L^2$
that gives an effective total spring constant of the bound leg.  In
the strained telemark stance of the waiting state, when both legs are
bound and Po, and the L leg is bent backward from $\theta_c =
60^\circ$ to about $\theta_c^\prime = 120^\circ$, $\delta z \approx L$
and the effective spring is loaded with a mechanical energy of
$E_\text{wait} = k \delta z^2/2 = {\cal T}/2\beta$.  This is
essentially the energy necessary for the power stroke (Pr to Po)
transition that loads the spring.  For $l_p = 310$ nm and $\nu_c =
184$ we have ${\cal T} = 23.2$, $k = 0.078$ pN/nm, and $E_\text{wait}
= 11.6$ $k_BT$.  If the total energy available from ATP hydrolysis is
$\approx 24$ $k_BT$, then this corresponds to a thermodynamic
efficiency of nearly $50\%$, similar to earlier estimates for myosins
V~\cite{Lan05} and II~\cite{Barclay98}.  The tension in the waiting
state, associated with this stored mechanical energy, is
$F_\text{wait} = k \delta z = 2.7$ pN.

  Myosin II offers an interesting point of comparison in terms of
  mechanical compliance.  The stiffness $k$ of its S1 domain is a key
  parameter in the swinging crossbridge model of muscle
  contraction, with a range $k \approx 1-3$ pN/nm inferred from
  experimental measurements~\cite{Howard96,Decostre05,Lewalle08}, an
  order of magnitude higher than our MyoV value above.  The key factor
  underlying this difference is the length of the lever arm, with
  myosin II having an $L$ about 1/3 that of MyoV.  If one assumes that
  beyond this difference the other structural factors ($l_p$ and
  $\nu_c$) are similar between these two systems, then one can use our
  structural model with $l_p = 310$ nm, $\nu_c=184$, and $L=12$ nm to
  predict a myosin II stiffness of $k = 1.5$ pN/nm, which compares
  well with the experimental range.

\vspace{2em}

\tocless\section{Conclusion}

In conclusion, we have proposed a model of MyoV dynamics, based on the
polymeric nature of the lever arms and the probability distribution of
their fluctuations during the diffusive search for actin binding
sites.  Using only three experimentally unknown parameters, our theory
quantitatively captures many experimental outcomes, such as the time
dependence of the mean trajectory of the detached head and the force
dependence of the probability ratio of forward to backward stepping.
The theory, which allows us to explore the robustness of stepping to
variations in the design of MyoV, also yields testable predictions for
novel quantities, like the probabilities of foot stomping as a
function of load.  Though the unidirectionality of the motor and the
stall force magnitude exhibit tolerance to variation in the structural
parameters, the theory reveals constraints on the persistence length
of the lever arms and power stroke bias.  In the context of
processive motors within the myosin superfamily, MyoV has the simplest
lever arm structure, which can be approximated well by a stiff
polymer.  Myosins VI and X have evolved qualitatively different lever
arms, consisting of both stiff and flexible segments~\cite{Sun11}.
The underlying theoretical ideas in our description of MyoV are quite
general, and it will be interesting to extend them in the future to
more complex geometries.  How do the structural constraints change in
a motor with heterogeneous persistence length, and can such an
approach help resolve the competing hypotheses for the conformation of
the myosin VI lever arm~\cite{Spink08,Mukherjea09,Zhang10}? 

From a broader perspective the approach we have developed is also
applicable in other motor systems, such as dynein and kinesin,
provided the structural elements generating the power stroke can
be modeled as suitable polymer chains. In addition, there are
potential applications to other biological systems that transmit or
generate force, such as mictotubules and cytoskeletal structures.

\vspace{2em}

\tocless\section{Materials and Methods}

\tocless\subsection{First passage times to binding sites}  The derivation of
Eq.~\eqref{e1} for the mean first passage times $t_\text{fp}^\pm$ is
shown in detail in the SI.  The underlying approach is based on the
renewal method for first passage problems~\cite{vanKampen}, which in
the polymer context is equivalent to the Wilemski-Fixman theory for
diffusion-controlled reactions~\cite{Wilemski1974a}.  For analytical
tractability we ignore excluded volume interactions, which would
likely lead to a small decrease in the first-passage times, but not
change the overall order of magnitude.  Strictly speaking,
$t_\text{fp}^\pm$ depends on the initial configuration of the polymer,
but for MyoV dynamics $t_\text{fp}^\pm \gg t_\text{r}$, the
relaxation time of the polymer to equilibrium.  Hence the memory of
the initial configuration is lost during the diffusive search, and the
expression for $t_\text{fp}^\pm$ in Eq.~\eqref{e1} is valid assuming
we do not start with the free end in the immediate vicinity of the
target.  When the latter condition is violated, for example after
failed binding attempts due to wrong head orientation, or immediately
following detachment from the actin, we assume fast relaxation to
equilibrium before the head has a chance to rebind.

\vspace{2em}

\tocless\subsection{Mean field theory for probability distribution of MyoV
  free end during diffusive search} The key physical quantity in
Eq.~\eqref{e1} that determines the average first passage time to a
binding site is ${\cal P}(\mb{r}_\pm)$, the equilibrium probability
density of finding the detached end of MyoV at $\mb{r}_\pm = \pm
\Delta \hmb{z}$.  For a structure of two semiflexible polymer legs,
with one leg bound at the origin, the free end-point vector $\mb{r} =
\mb{r}_f + \mb{r}_b$, where $\mb{r}_{f/b}$ is the end-to-end vector of
the free/bound leg.  The distribution ${\cal P}(\mb{r})$ is a
convolution of the individual leg distributions ${\cal
  P}_{f/b}(\mb{r}_{f/b})$,
\begin{equation}\label{m0}
{\cal P}(\mb{r}) = \int d\mb{r}_b\int d\mb{r}_f\, {\cal P}_b(\mb{r}_b) {\cal P}_f(\mb{r}_f) \delta(\mb{r}-\mb{r}_b-\mb{r}_f).
\end{equation}
  There is no exact closed form expression
for the end-to-end distribution of a semiflexible polymer, though
moments of the distribution can be calculated
analytically~\cite{Kratky49,Saito67}.  For the free leg,
  which is not under tension, an earlier mean-field
  theory~\cite{Hyeon06} gives a useful approximation,
\begin{equation}\label{m1}
{\cal P}_f (\mb{r}_f) = A_f \xi_f^{-9/2} \exp\left(-\frac{3\kappa}{4 \xi_f}\right),
\end{equation}
where $\kappa = L/l_p$, $\xi_f = 1-r_f^2/L^2$, and $A_f$ is a
normalization constant,
\begin{equation}\label{m2}
A_f = \frac{9 \sqrt{3} e^{3 \kappa/4} \kappa^{7/2}}{8 \pi ^{3/2} L^3 \left(3 \kappa^2+12 \kappa+20\right)}.
\end{equation}
As shown in the SI, this mean-field approach can be generalized to
include the end-tangent constraint and load force in the bound leg
case, yielding
\begin{equation}\label{m3}
{\cal P}_b (\mb{r}_b) = A_b \xi_b^{-9/2} \exp\left(-\frac{3\kappa}{4 \xi_b} + {\cal T}^\prime \hat{\mb{u}}^\prime_c \cdot \hat{\mb{r}}_b\right),
\end{equation}
where $\xi_b = 1-r_b^2/L^2$, $\hat{\mb{r}}_b= \mb{r}_b/r_b$, ${\cal T}^\prime = \sqrt{({\cal T}^\prime_x)^2+({\cal T}^\prime_z)^2}$ and
\begin{equation}\label{m4}
\begin{split}
{\cal T}^\prime_x &= {\cal T} \sin\theta_c + \beta F L \sin\theta_F,\\
{\cal T}^\prime_z &= {\cal T} \cos\theta_c - \beta F L \cos\theta_F.
\end{split}
\end{equation}
The power stroke effectiveness parameter ${\cal T}$ is defined in
Eq.~\eqref{e3}.  The direction $\hat{\mb{u}}^\prime_c =
\sin\theta_c^\prime\hmb{x} + \cos\theta_c^\prime \hmb{z}$, with an
angle $\theta_c^\prime$ from the $\hmb{z}$ axis given by
\begin{equation}\label{m5}
\theta_c^\prime = \theta_c + \tan^{-1}\left(\frac{\beta FL \sin (\theta_c+\theta_F)}{{\cal T}-\beta F L \cos (\theta_c+\theta_F)} \right).
\end{equation}
In the limit of large $l_p$, the vector $\hat{\mb{u}}^\prime_c$ is
approximately the average orientation of the bound leg, reflecting the
combined influence of the load force $F$ and the end-tangent
constraint $\nu_c$.  In the case of a backward force ($\theta_F =0$),
we can invert Eq.~\eqref{m5} to find the force $F$ required on average
to maintain an orientation $\theta_c^\prime > \theta_c$, as shown in
Eq.~\eqref{e12}.  In both the free and bound leg cases the analytical
distributions ${\cal P}_{f/b}(\mb{r}_{f/b})$ have excellent agreement
with the exactly known moments.  Carrying out the convolution in
Eq.~\eqref{m0}, we arrive at a final expression for ${\cal
  P}(\mb{r}_\pm)$ in the stiff regime ($l_p \gg L$):
\begin{equation}\label{m4}
\begin{split}
&{\cal P}(\mb{r}_\pm)\\
&\approx \frac{(3 \kappa  (7 \kappa +20)+200) {\cal T}^\prime}{1600 \pi  L^2 \Delta \sinh {\cal T}^\prime} I_0\left({\cal T}^\prime_x \sqrt{1 - (\Delta/2L)^2}\right)   e^{\pm \frac{{\cal T}^\prime_z \Delta}{2 L}},
\end{split}
\end{equation}
where $I_0(x)$ is the zeroth-order modified Bessel function of the first
kind.

\vspace{2em}

\SkipTocEntry\begin{acknowledgments}
  M.H. was a Ruth L. Kirschstein National Research Service
  postdoctoral fellow, supported by a grant from the National
  Institute of General Medical Sciences (1 F32 GM 97756-1). D.T. was
  supported by a grant from the National Science Foundation (CHE
  09-10433) and the National Institutes of Health (GM 089685).
\end{acknowledgments}

\begin{widetext}

\renewcommand{\theequation}{S\arabic{equation}}
\renewcommand{\thefigure}{S\arabic{figure}}
\renewcommand{\thetable}{S\arabic{table}}
\setcounter{equation}{0}
\setcounter{figure}{0}
\setcounter{section}{0}

\newpage

\baselineskip=20.74pt
\begin{center}
{\Large\bf Supplementary information for ``Design principles governing the motility of myosin V''}\\[1em]
\baselineskip=17.28pt
{\large Michael Hinczewski, Riina Tehver and D. Thirumalai}
\end{center}
\baselineskip=14.4pt


In this SI we provide the details of the theory that nearly
quantitatively explains the complex kinetic pathways in the stepping
dynamics of myosin V (MyoV). Because the SI is long, containing
technical details of the calculations, we begin with a collection of
the most important equations, which were used to make the predictions
described in the main text. The subsequent sections describe the
details leading to these equations.

\section{Summary of key equations for myosin V dynamics}

\renewcommand{\arraystretch}{1.5}
\[
\begin{array}{lr}
\text{\underline{\em First passage and binding}} & \\[0.25em]
t^\pm_\text{fp} = \frac{1}{4\pi a D_\text{h} {\cal P}(\mb{r}_\pm)}, \qquad
t_\text{Tb} = t_\text{h} + \frac{t_\text{fp}^+}{1+b\alpha}, \qquad t_\text{Lb} = \frac{t_\text{fp}^+}{b+\alpha}, \qquad \alpha = t_\text{fp}^+/t_\text{fp}^- &  \text{\eqref{f9}}, \text{\eqref{f10d}, \eqref{f12d}}\\[1em]
\text{\underline{\em Kinetic pathway probabilities}} & \\[0.25em]
{\cal P}_\text{f} = \frac{g}{1+g}\frac{t_\text{d1}^2}{(1+b\alpha)(t_\text{d1}+t_\text{h})(t_\text{d1} + t_\text{Tb} - t_\text{h})}, \qquad {\cal P}_\text{Ts} =b\alpha {\cal P}_\text{f} & \text{\eqref{f10}, \eqref{f11}}\\
{\cal P}_\text{Ls} = \frac{1}{1+g}\frac{b t_\text{d1}}{(b+\alpha)(t_\text{d1} + t_\text{Lb})}, \qquad {\cal P}_\text{b} = b^{-1}\alpha {\cal P}_\text{Ls} & \text{\eqref{f12}, \eqref{f13}}\\[1em]
\text{\underline{\em Average step shape}} &\\[0.25em]
{\cal P}^+_\text{Tb}(t) =  \frac{t_\text{h}\left(1-e^{-t/t_\text{h}}\right) - t_\text{fp}^+(1+b\alpha)^{-1} \left(1-e^{-t(1+b\alpha)/t_\text{fp}^+}\right)}{t_\text{h}(1+b\alpha)-t_\text{fp}^+}, \qquad {\cal P}^-_\text{Tb}(t) =b \alpha {\cal P}^+_\text{Tb}(t) & \text{\eqref{f14}, \eqref{f15}}\\
\langle \delta z(t) \rangle = (\mu_z+\Delta)(1-{\cal P}^+_\text{Tb}(t)-{\cal P}^-_\text{Tb}(t))(1-e^{-t/t_\text{r}})+ 2\Delta {\cal P}^+_\text{Tb}(t) & \text{\eqref{f16}}\\
\mu_z = l_p\left(1-e^{-L/l_p}
\right) \left(\coth\nu_c - \nu_c^{-1}\right)\cos\theta_c & \text{\eqref{f17}}\\[1em]
\text{\underline{\em Mean run length and velocity}} &\\[0.25em]
z_\text{run} = v_\text{run} t_\text{run}, \qquad v_\text{run} \approx \frac{\Delta}{t_\text{d1}}\left( \frac{1}{1+b\alpha} - \frac{\alpha}{g(b+\alpha)}\right), \qquad t_\text{run} \approx \frac{g t_\text{d1}^2}{t_\text{Lb}+ g t_\text{Tb}} & \text{\eqref{f18}-\eqref{f21}}\\[1em]
\text{\underline{\em Equilibrium end-point probability distribution}} &\\[0.25em]
{\cal T} \approx 1+ \frac{20 \nu_c}{20+7\kappa \nu_c}, \qquad {\cal T}^\prime = \sqrt{({\cal T}^\prime_x)^2+({\cal T}^\prime_z)^2} & \text{\eqref{s13}, \eqref{s14}}\\
{\cal T}^\prime_x = {\cal T} \sin\theta_c + \beta F L
\sin\theta_F, \qquad {\cal T}^\prime_z = {\cal T} \cos\theta_c - \beta
F L \cos\theta_F & \text{\eqref{s14}}\\
{\cal P}(\mb{r}_\pm) \approx \frac{(3 \kappa  (7 \kappa +20)+200) {\cal T}^\prime}{1600 \pi  L^2 \Delta \sinh {\cal T}^\prime} I_0\left({\cal T}^\prime_x \sqrt{1- \frac{\Delta^2}{4 L^2}}  \right)   e^{\pm \frac{{\cal T}^\prime_z \Delta}{2 L}} & \text{\eqref{s23}}\\[1em]
\text{\underline{\em Stall force}} &\\[0.25em]
F_\text{stall} = \frac{k_B T}{\cos \theta_F}\left(\frac{{\cal T}}{L} \cos\theta_c + \frac{1}{\Delta} \log  \frac{g-1+\sqrt{(g-1)^2 +4gb^2}}{2 b}\right) & \text{\eqref{st4}}\\[1em]
\end{array}
\]

\section{First passage times, binding probabilities, and experimental observables}
 
\noindent {\bf Mean first passage time to a target site}. After the
detachment of one of the myosin V (MyoV) heads from the polar actin
tracks, there are two potential actin target sites where the head
could rebind, at positions $\mb{r}_\pm = \pm \Delta \hmb{z}$ (see
Fig.~1B in the main text).  The axis $\hmb{z}$ is oriented from the
minus to plus end of the actin filament, so we denote $\mb{r}_+$ and
$\mb{r}_-$ as the forward and backward target sites respectively.
Before dealing with the full complexity of the diffusive search and
binding for multiple targets (with binding probabilities dependent on
the head chemical state), we solve a simpler problem: what is the mean
first passage time for the free end of MyoV to reach a sphere of
radius $a$ around one of the target sites, for example $\mb{r}_+$?
(The derivation below will hold analogously for $\mb{r}_-$, with the
$+$ superscripts and subscripts replaced by $-$.)

Let $f_\text{fp}(\mb{r},\mb{r}^\prime;t)$ be the distribution of
first-passage times for the free end to go from an initial position
$\mb{r}$ to some final position $\mb{r}^\prime$.  Using the renewal
approach~\cite{vanKampen}, the first-passage time distribution can be
related to the Green's function $G(\mb{r},\mb{r}^\prime;t)$ describing
the probability of diffusing from $\mb{r}$ to $\mb{r}^\prime$ in time
$t$.  Choose a final position on a sphere of radius $a$ around the
target site $\mb{r}_+$, so $\mb{r}^\prime = \mb{r}_+ + a \hmb{e}$,
where $\hmb{e}$ is any unit vector.  The renewal approach relates
$f_\text{fp}$ and $G$ through the integral equation,
\begin{equation}\label{f2}
G(\mb{r},\mb{r}_++a\hmb{e};t) = \int_0^t dt^\prime\,\int
a^2 d\hmb{e}^\prime
f_\text{fp}(\mb{r},\mb{r}_++a\hmb{e}^\prime;t^\prime)
G(\mb{r}_++a\hmb{e}^\prime,\mb{r}_++a\hmb{e};t-t^\prime).
\end{equation}
The physical meaning of the equation above is that the Green's function for
going from $\mb{r}$ to a particular point $\mb{r}_++a\hmb{e}$ on the
target sphere consists of paths that make first-passage at some point
$\mb{r}_++a\hmb{e}^\prime$ on the target sphere at time $t^\prime\le
t$, and then diffuse from $\mb{r}_++a\hmb{e}^\prime$ to
$\mb{r}_++a\hmb{e}$ in time $t-t^\prime$.  Since
Eq.~\eqref{f2} is difficult to solve analytically, we make three
simplifications, motivated by the observation that the capture radius $a$ is
small compared to all other length scales in the problem: (i) we
approximately average over all final positions on the target sphere,
replacing $\mb{r}_++a\hmb{e}$ with $\mb{r}_+$ on both sides of
Eq.~\eqref{f2}; (ii) we assume
$f(\mb{r},\mb{r}_++a\hmb{e}^\prime;t^\prime)$ does not vary
appreciably with $\hmb{e}^\prime$, so it can be replaced by
$f^+_\text{fp}(\mb{r};t^\prime)/4\pi a^2$, where
$f^+_\text{fp}(\mb{r};t^\prime)$ is the first passage time
distribution for reaching any point on a target sphere of radius $a$
around $\mb{r}_+$, starting from $\mb{r}$; (iii) the Green's function on the right-hand side
of Eq.~\eqref{f2} will not depend significantly on the specific unit
vector $\hmb{e}^\prime$ defining the starting position, so we replace
$\hmb{e}^\prime$ in the argument of the Green's function by a fixed
unit vector $\hmb{z}$.  With these approximations, Eq.~\eqref{f2}
becomes:
\begin{equation}\label{f3}
G(\mb{r},\mb{r}_+;t) \approx \int_0^t dt^\prime\,f^+_\text{fp}(\mb{r};t^\prime) G(\mb{r}_++a\hmb{z},\mb{r}_+;t-t^\prime).
\end{equation}
The above renewal equation can be solved by Laplace transforming both
sides to yield:
\begin{equation}\label{f4}
\tilde{f}^+_\text{fp}(\mb{r};s) \approx \frac{\tilde{G}(\mb{r},\mb{r}_+;s)}{
\tilde{G}(\mb{r}_++a\hmb{z},\mb{r}_+;s)},
\end{equation}
where $\tilde{f}^+_\text{fp}$ and $\tilde{G}$ are Laplace-transformed
functions. For example, $\tilde{G}(\mb{r},\mb{r}^\prime;s) =
\int_0^\infty dt\,e^{-s t}G(\mb{r},\mb{r}^\prime;t)$, and a similar
equation holds for $\tilde{f}^+_\text{fp}$.  The derivative of
$\tilde{f}$ with respect to $s$ at $s=0$ is related to the mean
first-passage time $t^+_\text{fp}(\mb{r})$ to arrive at the target
sphere of radius $a$ around $\mb{r}_+$:
\begin{equation}\label{f5}
-\left.\frac{\partial}{\partial s}\tilde{f}^+_\text{fp}(\mb{r};s)\right|_{s=0} = \int_0^\infty dt\, t f^+_\text{fp}(\mb{r};t) = t^+_\text{fp}(\mb{r})
\end{equation}

We can simplify Eq.~\eqref{f4} by taking advantage of time scale
separation in the system.  For $t \gg t_\text{r}$, the relaxation time of the
two-legged polymer, the Green's function for going from an initial to
final position approaches the equilibrium probability distribution of
finding the free end at the final position, $G(\mb{r},\mb{r}^\prime;t)
\to {\cal P}(\mb{r}^\prime)$ as $t \to \infty$.  In Laplace space,
this implies that the Green's function can be decomposed into two
contributions,
\begin{equation}\label{f6}
\begin{split}
\tilde{G}(\mb{r},\mb{r}^\prime;s) &\approx \int_0^{t_\text{r}} dt\,e^{-st} G(\mb{r},\mb{r}^\prime;t) + (s^{-1} - t_\text{r}) {\cal P}(\mb{r}^\prime)\\
& \equiv \tilde{G}_0(\mb{r},\mb{r}^\prime;s) + (s^{-1}-t_\text{r}){\cal P}(\mb{r}^\prime).
\end{split}
\end{equation}
For $\tilde G(\mb{r},\mb{r}_+;s)$ in the numerator of Eq.~\eqref{f4},
we assume the initial $\mb{r}$ is not in the immediate vicinity of the
target $\mb{r}_+$ (which is generally the case for MyoV
diffusive search), so the time to reach the target will be much larger
than the relaxation time $t_\text{r}$.  Hence $\tilde G_0(\mb{r},\mb{r}_+;s)$ will
be negligible, because $G(\mb{r},\mb{r}_+;t)$ is near zero on the time
scale $t < t_\text{r}$.  Thus we can approximate the numerator of
Eq.~\eqref{f4} as:
\begin{equation}\label{f7}
\tilde G(\mb{r},\mb{r}_+;s) \approx (s^{-1}-t_\text{r}){\cal P}(\mb{r}_+).
\end{equation}
For the denominator of Eq.~\eqref{f4},
$\tilde{G}(\mb{r}_++a\hmb{z},\mb{r}_+;s)$, the situation is more
complicated, because the initial and final positions are separated by
a small distance $a$, and hence there will be contributions to $\tilde
G_0$ at short times.  In the limit $a \to 0$, the paths between
$\mb{r}_++a\hmb{z}$ and $\mb{r}_+$ involve only a fast, microscopic
rearrangment of the free end, without significant configurational
changes in the rest of the structure.  If we model the free end as a
particle with diffusion constant $D$, the Green's function in the
short time limit can be approximated as~\cite{vanKampen}:
\begin{equation}\label{f7b}
G(\mb{r}_++a\hmb{z},\mb{r}_+;t)
\approx (4\pi D t)^{-3/2} \exp(-a^2/(4Dt)).
\end{equation}

Substituting Eq.~\eqref{f7b} into the integral for $\tilde G_0$, we get an
expression for the denominator,
\begin{equation}\label{f8}
\begin{split}
\tilde G(\mb{r}_++a\hmb{z},\mb{r}_+;s) &\approx \frac{t_\text{a}}{4\pi a^3}\text{erfc}\left(\frac{1}{2}\sqrt{\frac{t_\text{a}}{t_\text{r}}}\right) + (s^{-1}-t_\text{r}){\cal P}(\mb{r}_+)\\
&\approx \frac{t_\text{a}}{4\pi a^3} + (s^{-1}-t_\text{r}){\cal P}(\mb{r}_+),
\end{split}
\end{equation}
where $t_\text{a} = a^2/D$ is a microscopic time scale describing how
long it takes a particle of diffusivity $D$ to move a distance $a$.
The second approximation in Eq.~\eqref{f8} assumes $t_\text{a} \ll
t_\text{r}$, which is justified by a simple calculation: let us set
$D= D_\text{h}$, where $D_\text{h} = 5.7 \times 10^{-7}$ cm$^{2}$/s is
the diffusion constant of the MyoV head, as derived from the PDB
structure 1W8J~\cite{Coureux04} using the program
HYDROPRO~\cite{Ortega11}.  For $a=1$ nm, the resulting microscopic
time scale $t_\text{a} = 18$ ns, which is significantly smaller than
the relaxation time $t_\text{r} \sim {\cal O}(1\:\mu\text{s})$ of the
entire structure (see the next subsection for estimates of
$t_\text{r}$).  

Using Eqs.~\eqref{f7} and \eqref{f8} in
Eq.~\eqref{f4}, and then evaluating the derivative in Eq.~\eqref{f5},
we obtain the final approximate expression for the mean first passage
time:
\begin{equation}\label{f9}
t^+_\text{fp} = \frac{1}{4\pi a D_\text{h} {\cal P}(\mb{r}_+)}.
\end{equation}
We have dropped the $\mb{r}$ dependence in the notation for
$t^+_\text{fp}(\mb{r})$, since the first passage time result is
independent of the initial position $\mb{r}$.  This reflects the
underlying assumption that the configurational relaxation time $t_\text{r}
\ll t^+_\text{fp}$, so the free end loses memory of its initial
position during the long diffusive search.  An analogous result holds
for the mean first passage time $t^-_\text{fp}$ to the backward target
site, with $\mb{r}_+$ replaced by $\mb{r}_-$ in Eq.~\eqref{f9}. A result similar in spirit to 
Eq.~\eqref{f9} but without the benefit of derivation, was conjectured earlier~\cite{Guo95BP}.  

\begin{figure}[t]
\centering\includegraphics*[width=0.6\textwidth]{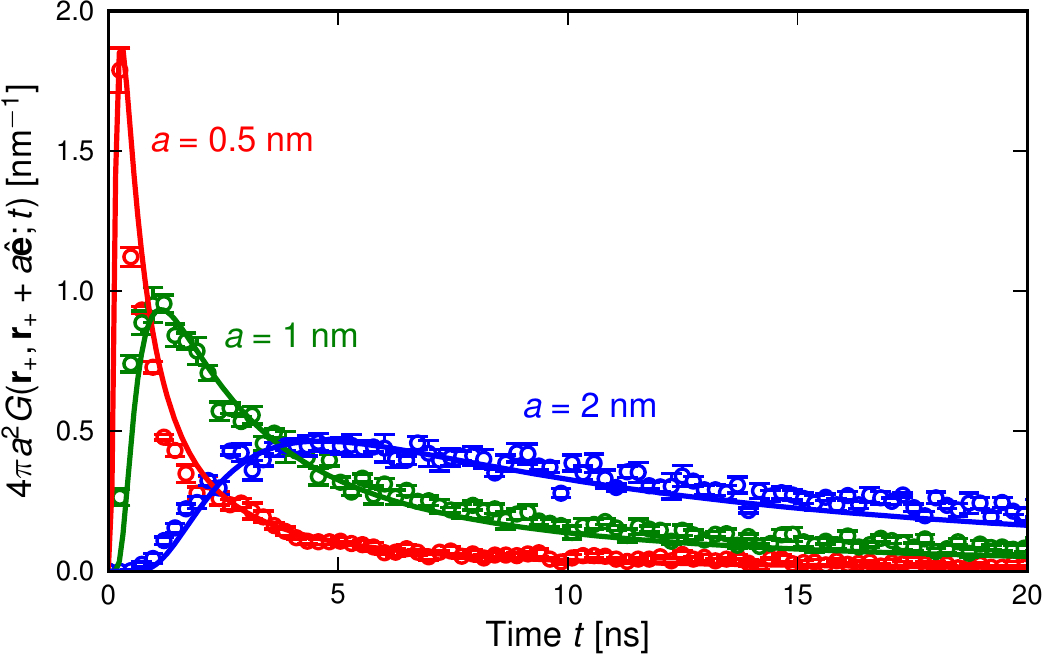}
\caption{Brownian dynamics simulation results (circles) for the
  Green's functions of the end-point of the two-legged MyoV
  structure, with $\nu_c = 184$, $l_p = 310$ nm, and $\theta_c =
  60^\circ$.  The plot shows $4 \pi a^2
  G(\mb{r}_+,\mb{r}_++a\hmb{e};t)$ as a function of time $t$, where
  $G(\mb{r}_+,\mb{r}_++a\hmb{e};t)$ is the probability of diffusing a
  distance $a$ from $\mb{r}_+$ to some point $\mb{r}_++a\hmb{e}$, with
  $|\hmb{e}|=1$.  Results for three different values of $a$ are displayed:
  $a= 0.5$ nm (red); $a=1$ nm (green); $a=2$ nm (blue).  Error bars
  denote standard error for the simulation-derived values.  For
  comparison, the solid curves represent the expression $4\pi a^2
  (4\pi D t)^{-3/2} \exp(-a^2/(4Dt))$, the right-hand-side of
  Eq.~\eqref{f7b} multiplied by $4\pi a^2$, with a best-fit value of
  $D = 1.4 \pm 0.1 \times 10^{-6}$ cm$^2$/s.}\label{gr}
\end{figure}

In order to validate the approximation underlying Eq.~\eqref{f7b}, we
performed Brownian dynamics simulations on a bead-spring semiflexible
polymer model of two-legged MyoV (further details are in the next
section, ``Relaxation times'').  By generating many individual
trajectories of the detached polymer end-point diffusing a small
distance $a$ from $\mb{r}_+$ to some point $\mb{r}_+ + a\hmb{e}$, we
numerically reconstruct the corresponding Green's function
(Fig.~\ref{gr}).  The excellent fit of the assumed form in
Eq.~\eqref{f7b} for several values of $a$ to the numerical results
justifies the approximation.  

\vspace{1em}

\noindent {\bf Relaxation times}.  To estimate the relaxation
$t_\text{r}$ of the two-legged MyoV structure, we performed Brownian
dynamics~\cite{Ermak78} simulations of a bead-spring semiflexible
polymer model.  Each leg consists of 17 beads of diameter $d=2$ nm,
with an additional bead at the flexible joint between the legs.  The
beads are connected through harmonic springs of stiffness 200
$k_BT$/nm$^2$ and each leg has a bending elasticity described by a
persistence length $l_p = 50-400$ nm.  Initially the end beads are
fixed at the two binding sites.  The end-tangent of the leading leg
(the unit vector oriented between the centers of the first two beads)
is subject to a harmonic constraint of strength $k_BT \nu_c$ along
$\hmb{u}_c$ (at an angle $\theta_c = 60^\circ$ from the actin
filament), with $\nu_c = 50-180$.  The joint between the legs is
subject to a backward load force of $F$.  The beads are coupled
hydrodynamically through the Rotne-Prager tensor~\cite{Rotne69}, and
their positions evolve in time numerically according to the Langevin
equation.  Each simulation lasts 12 $\mu$s, where during the first 2.4
$\mu$s both end beads are bound, and during the remaining time the
trailing leg end bead is allowed to diffuse freely.  By averaging a
large number of individual simulations (1000-1250 runs for each
distinct parameter set of $l_p$ and $\nu_c$) we can extract the mean
time $t_\text{r}$ for the $z$-axis position of the trailing leg end
bead to reach equilibrium after detachment.

\begin{figure}[t]
\centering\includegraphics*[width=0.9\textwidth]{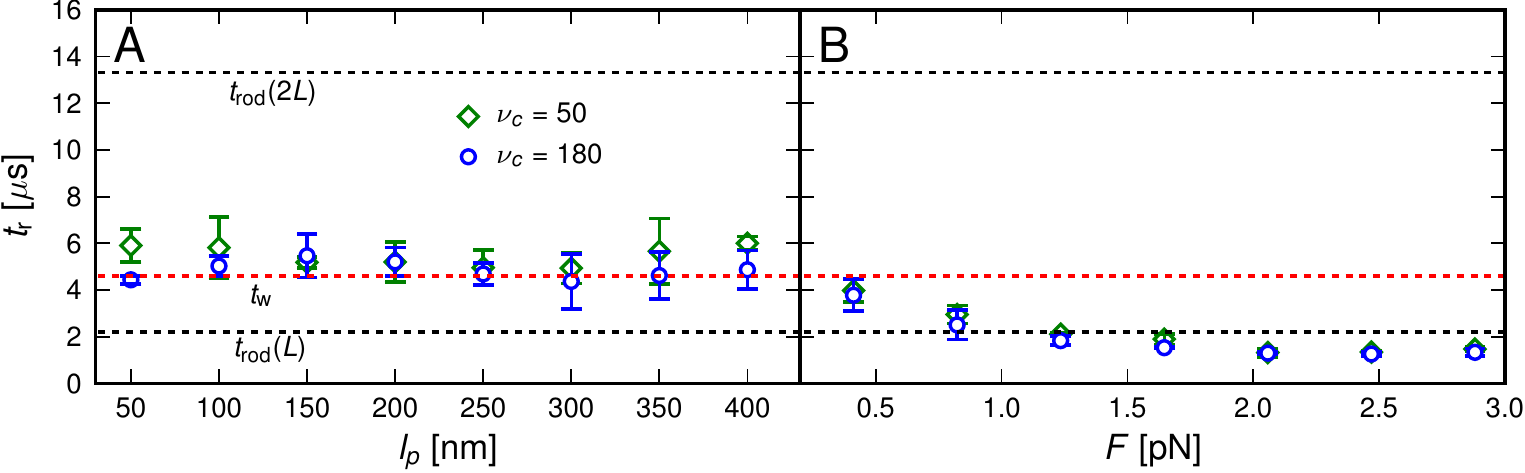}
\caption{Relaxation times $t_\text{r}$ for the trailing end-point of
  the two-legged MyoV structure to reach equilibrium after
  detachment, calculated from Brownian dynamics simulations.  Results
  are shown at two different strengths $\nu_c=50,\:180$ of the bound
  leg power stroke constraint, with $\theta_c=60^\circ$.  A)
  $t_\text{r}$ at zero load as a function of leg persistence length
  $l_p$. B) $t_\text{r}$ at $l_p = 310$ nm as a function of backward
  load force $F$.  For comparison three analytically estimated
  rotational diffusion times are shown as horizontal dashed lines:
  $t_\text{rod}(L)$ and $t_\text{rod}(2L)$ (black) [Eq.~\eqref{tr1}], for a
  rigid rod of length $L$ and $2L$ respectively, and $t_\text{w}(L)$
  (red) [Eq.~\eqref{tr2}] for two rigid rods of length $L$ connected at a
  flexible hinge.}\label{tr}
\end{figure}

Fig.~\ref{tr} shows the resulting values of $t_\text{r}$ for $\nu_c =
50$ and $180$, with panel A plotting $t_\text{r}$ as a function of
$l_p$, and panel B as a function of backward load force $F$ at $l_p =
310$ nm.  In the absence of load, $t_\text{r} \approx 5$ $\mu$s for
both values of $\nu_c$ over the entire plotted range of $l_p$
(corresponding to the semiflexible regime $l_p > L$).  Since
relaxation of MyoV requires a rotational reorientation of a
stiff, two-legged structure (with each leg of contour length $L=35$
nm), we expect that $t_\text{r}$ should fall in the range between the
rotational diffusion time $t_\text{rod}(L)$ of a rigid rod of length
$L$, and $t_\text{rod}(2L)$, the time for a rigid rod of length
$2L$.  Analytically, $t_\text{rod}(L)$ can be approximated
as~\cite{DoiEdwards}:
\begin{equation}\label{tr1}
t_\text{rod}(L) = \frac{\pi \eta L^3}{3 \ln (L/2d)},
\end{equation}
where $\eta$ is the viscosity of water.  The resulting rotational
diffusion times $t_\text{rod}(L) \approx 2.2$ $\mu$s and
$t_\text{rod}(2L) \approx 13.3$ $\mu$s are marked as black dashed
lines in Fig.~\ref{tr}A, which establishes that $t_\text{rod}(L) <
t_\text{r} < t_\text{rod}(2L)$.  A more precise analytical comparison
can be made with the rotational diffusion time $t_\text{w}$ of a
structure consisting of two rigid rods of length $L$ connected by a
flexible hinge, which has been estimated by Wegener~\cite{Wegener80}:
\begin{equation}\label{tr2}
t_\text{w}(L) \approx 1.79 \frac{\pi \eta L^3}{\ln (2L/d)}.
\end{equation}
The resulting value $t_\text{w}(L) = 4.6$ $\mu$s, marked as a red
dashed line in Fig.~\ref{tr}A, is in good agreement with the
simulation results.  With a load force $F$ applied to MyoV, the
equilibrium position of the end-point after detachment is shifted
closer to the initial binding site.  As a result, the relaxation times
become shorter, as seen in Fig.~\ref{tr}B.  In all cases, $t_\text{r}$
is at least two orders of magnitude smaller than the typical first
passage times to the binding site, which is consistent with the
approximation used to derive Eq.~\eqref{f9}.

\vspace{1em}

\noindent {\bf Binding probabilities}.  When MyoV is in the waiting
state, with both heads bound to ADP and strongly associated with
actin, we can have one of two scenarios for initiating a diffusive
search: (i) ADP is released from the trailing head (TH), quickly
replaced by ATP, leading to the dissociation of the TH from actin.
This detachment through ADP release / ATP binding has a overall rate
$t_\text{d1}^{-1}$; (ii) less frequently, the leading head (LH)
detaches without ADP release, which occurs at a rate $t_\text{d2}^{-1}
\ll t_\text{d1}^{-1}$.  The gating parameter $g =
t_\text{d2}/t_\text{d1} \gg 1$ describes the probabilities of the two
scenarios occurring, which are $g(1+g)^{-1}$ for (i) and $(1+g)^{-1}$
for (ii).

Let us consider scenario (i), which can lead either to a forward step,
if the TH rebinds to $\mb{r}_+$, or a trailing foot stomp, if the TH
binds to $\mb{r}_-$.  Denote the probabilities of these two binding
events as ${\cal P}_\text{f}$ and ${\cal P}_\text{Ts}$.  For the TH to
bind to actin, three conditions must be fulfilled: (a) the TH must
hydrolyze ATP, which occurs at a rate $t_\text{h}^{-1}$; (b)
subsequently, the TH must reach the capture radius $a$ of one of the
binding sites.  For $\mb{r}_+$, it reaches the capture radius with
rate $(t_\text{fp}^+)^{-1}$, and then binds.  For $\mb{r}_-$, it
reaches the capture radius with rate $(t_\text{fp}^-)^{-1}$, but
binding will only occur with probability $b$, reflecting the penalty
for wrong head orientation after the recovery stroke.  Thus, the
effective rate of capture at the backward site is
$b(t_\text{fp}^-)^{-1}$ with $b \ll 1$ (see Table 1 in the main text);
(c) during the entire diffusive search, the LH must not detach from
actin, or the entire MyoV structure will dissociate from the filament
and the run is terminated.  The detachment rate, assumed to be
ATP-independent, is given by $t_\text{d1}^{-1}$.  Requirements (a) and
(b) by themselves, and the assumption that individual events are
Poisson distributed, lead to probability distributions
$f_\text{Tb}^\pm(t)$ for the TH binding time to the $\mb{r}_\pm$
target sites:
\begin{equation}\label{f10b}
\begin{split}
f^+_\text{Tb}(t) &= \int_0^{t} dt^\prime\,t_\text{h}^{-1} e^{-t^\prime/t_\text{h}} (t^+_\text{fp})^{-1} e^{-(t-t^\prime)\left[ (t^+_\text{fp})^{-1} +b (t^-_\text{fp})^{-1}\right]}\\
&=\frac{e^{-t/t_\text{h}}-e^{-t (1+ b\alpha)/t^+_\text{fp}}}{t_\text{h}(1+b \alpha)-t^+_\text{fp}},
\end{split}
\end{equation}
\begin{equation}\label{f10c}
\begin{split}
f^-_\text{Tb}(t) &= \int_0^{t} dt^\prime\,t_\text{h}^{-1} e^{-t^\prime/t_\text{h}} b(t^-_\text{fp})^{-1} e^{-(t-t^\prime)\left[ (t^+_\text{fp})^{-1} +b (t^-_\text{fp})^{-1}\right]}\\
&=b \alpha f^+_\text{Tb}(t),
\end{split}
\end{equation}
where $\alpha = t_\text{fp}^+ / t_\text{fp}^-$.  The integrals in
Eq.~\eqref{f10b} are convolutions of the probability that
hydrolysis occurs at some time $t^\prime$ and the probability of
subsequent capture at a target site after a time interval
$t-t^\prime$.  The average time to bind, $t_\text{Tb}$, is the same for
both sites:
\begin{equation}\label{f10d}
t_\text{Tb} = \frac{\int_0^\infty dt^\prime\, t^\prime
  f_\text{Tb}^+(t^\prime)}{\int_0^\infty dt^\prime f_\text{Tb}^+(t^\prime)} =
\frac{\int_0^\infty dt^\prime\, t^\prime
  f_\text{Tb}^-(t^\prime)}{\int_0^\infty dt^\prime f_\text{Tb}^-(t^\prime)} = t_\text{h} + \frac{t_\text{fp}^+}{1+b\alpha}.
\end{equation}

Using Eq.~\eqref{f10b} it is straightforward
to incorporate requirement (c) and derive the probabilities
${\cal P}_\text{f}$ and ${\cal P}_\text{Ts}$:
\begin{align}
{\cal P}_\text{f} &= \frac{g}{1+g}\int_0^\infty dt\, e^{-t/t_\text{
d1}} f^+_\text{Tb}(t)\nonumber\\
&= \frac{g}{1+g}\frac{t_\text{d1}^2}{(1+b\alpha)(t_\text{d1}+t_\text{h})(t_\text{d1} + t_\text{Tb} - t_\text{h})}.\label{f10}\\
{\cal P}_\text{Ts} &= \frac{g}{1+g}\int_0^\infty dt\, e^{-t/t_\text{d1}} f^-_\text{Tb}(t)\nonumber\\
&=b\alpha {\cal P}_\text{f}.\label{f11}
\end{align}

In scenario (ii), ATP hydrolysis is not required for rebinding, since
the detached LH retains ADP and is in a state that can strongly
associate with actin.  The head orientation is now favorable for
binding to the backward site, so the binding penalty exists for
$\mb{r}_+$ instead of $\mb{r}_-$.  The free LH can bind to
$\mb{r}_+$, an L foot stomp with probability ${\cal P}_\text{Ls}$, or
it can bind to $\mb{r}_-$, a backward step with probability ${\cal
  P}_\text{b}$.  The LH analogues to Eqs.~\eqref{f10b}-\eqref{f11}
can be obtained from these equations by the substitutions $t_\text{h}
=0$, $b(t^-_\text{fp})^{-1} \to (t^-_\text{fp})^{-1}$,
$(t^+_\text{fp})^{-1} \to b(t^+_\text{fp})^{-1}$.  The results are:
\begin{align}
f^+_\text{Lb}(t) &=b(t^+_\text{fp})^{-1} e^{-t(b+\alpha)/t_\text{fp}^+},\label{f12b}\\
f^-_\text{Lb}(t) &= b^{-1} \alpha f^+_\text{Lb}(t),\label{f12c}\\
t_\text{Lb} &= \frac{t_\text{fp}^+}{b+\alpha},\label{f12d}\\
{\cal P}_\text{Ls} &= \frac{1}{1+g}\frac{b t_\text{d1}}{(b+\alpha)(t_\text{d1} + t_\text{Lb})},\label{f12}\\
{\cal P}_\text{b} &= b^{-1}\alpha {\cal P}_\text{Ls}.\label{f13}
\end{align}
The final kinetic pathway, termination by complete dissociation from
actin, occurs when the diffusive search in any of the four pathways
above cannot be completed before the bound leg detaches.  The
termination probability is ${\cal P}_\text{t} = 1 - {\cal
  P}_\text{f}- {\cal P}_\text{Ts}-{\cal P}_\text{Ls}-{\cal P}_\text{b}$.

From Eqs.~\eqref{f10}, \eqref{f11}, \eqref{f12}, \eqref{f13}, one can
derive the pathway probability ratios shown in Eq.~(4) of the main
text.  The results for the ratios have been simplified under the
assumption that $t_\text{d1} \gg t_\text{Lb},\:t_\text{Tb}$, which is
generally valid.

\vspace{1em}

\noindent {\bf Average step shape}.  In order to compare with the
Dunn-Spudich experiment~\cite{Dunn07}, we will consider the average
step trajectory $\langle \delta z(t) \rangle$ of the TH along
$\hmb{z}$ after detachment from actin, where $\delta z(t) \equiv z(t)
- z(0)$, and the initial position is the backward binding site, $z(0)
= \hmb{z}\cdot \mb{r}_- = -\Delta$.  In the ensemble of all possible
trajectories at time $t$ after detachment (with at least one head
bound to actin), there will be two subpopulations: those trajectories
where the TH is still unbound, and those where the TH has
bound either to the backward site $\mb{r}_-$ or forward site
$\mb{r}_+$.  In this calculation we ignore the small fraction of
trajectories that lead to complete dissociation of the motor since
these are not counted as completed steps, and hence do not contribute
to the experimental measurement of $\langle \delta z(t) \rangle$.  The
fraction ${\cal P}^\pm_\text{Tb}(t)$ of TH trajectories that has
bound to $\mb{r}_\pm$ by time $t$ is:
\begin{align}
{\cal P}^+_\text{Tb}(t) &= \int_0^t dt^\prime\,f^+_\text{Tb}(t^\prime)\nonumber\\
&=  \frac{t_\text{h}\left(1-e^{-t/t_\text{h}}\right) - t_\text{fp}^+(1+b\alpha)^{-1} \left(1-e^{-t(1+b\alpha)/t_\text{fp}^+}\right)}{t_\text{h}(1+b\alpha)-t_\text{fp}^+},\label{f14}\\
{\cal P}^-_\text{Tb}(t) &= \int_0^t dt^\prime\,f^-_\text{Tb}(t^\prime) =b \alpha {\cal P}^+_\text{Tb}(t),\label{f15}
\end{align}
where $f^\pm_\text{Tb}(t)$ are the binding time distributions given by
Eq.~\eqref{f10b}-\eqref{f10c}.  The expression for the average step is
then:
\begin{equation}\label{f16}
\langle \delta z(t) \rangle = (\mu_z+\Delta)(1-{\cal P}^+_\text{Tb}(t)-{\cal P}^-_\text{Tb}(t))(1-e^{-t/t_\text{r}})+ 2\Delta {\cal P}^+_\text{Tb}(t). 
\end{equation}
The first term in Eq.~\eqref{f16} reflects the relaxation of the
unbound subpopulation over a characteristic time $t_\text{r}$ to the
average position of the free end along the $\hmb{z}$ axis, $\mu_z =
\langle \hmb{z}\cdot \mb{r}\rangle$, where $\mb{r}$ is the end-to-end
vector of MyoV, and the average is taken over the equilibrium
configurations of a two-legged polymer with one leg bound to the actin
filament, and the other leg free.  As described in the next section,
this average can be exactly derived, and is related to the structural
parameters of the system: the leg contour length $L$, the persistence
length $l_p$, strength of the end-tangent constraint $\nu_c$ at
the bound end, and the angle of the constraint direction $\theta_c$
relative to the $\hmb{z}$ axis.  The full expression for $\mu_z$ is:
\begin{equation}\label{f17}
\mu_z = l_p\left(1-e^{-L/l_p}
\right) \left(\coth\nu_c - \nu_c^{-1}\right)\cos\theta_c.
\end{equation}
For those interested in the derivation, $\mu_z =
\mu_\parallel^\text{exact} \cos\theta_c$, where
$\mu_\parallel^\text{exact}$ is given by Eq.~\eqref{s10} below.  The
value of the polymer relaxation time $t_\text{r} \approx 5$ $\mu$s,
as discussed above.  The second term in Eq.~\eqref{f16} is the
contribution of trajectories that have bound to $\mb{r}_+$, and hence
covered a distance $\delta z = 2\Delta$ along the filament axis.
Trajectories binding to the initial site $\mb{r}_-$ have a $\delta z
=0$, and so do not appear in Eq.~\eqref{f16}.

\vspace{1em}

\noindent {\bf Run length and velocity}.  If the termination probability during
each diffusive search is ${\cal P}_\text{t} = 1 - {\cal P}_\text{f}-
{\cal P}_\text{Ts}-{\cal P}_\text{Ls}-{\cal P}_\text{b}$, then the
mean number of searches during a run is $\sum_{n=1}^\infty n (1-{\cal P}_\text{t})^{n-1} {\cal P}_\text{t} = 1/{\cal P}_\text{t}$.  The
fraction of the searches within a run which lead to forward steps is
${\cal P}_\text{f}/(1-{\cal P}_\text{t})$, and the fraction which lead
to backward steps is ${\cal P}_\text{b}/(1-{\cal P}_\text{t})$.  The
mean run length, assuming step size $\Delta$, is given by:
\begin{equation}\label{f18}
z_\text{run} = \frac{\Delta ({\cal P}_\text{f}-{\cal P}_\text{b})}{{\cal P}_\text{t}(1-{\cal P}_\text{t})} \approx \frac{\Delta t_\text{d1} (\alpha(g-1)+b(g-\alpha^2))}{(b+\alpha)(1+b\alpha)(t_\text{Lb}+g t_\text{Tb})},
\end{equation}
where we have used the pathway probabilities from Eqs.~\eqref{f10},
\eqref{f11}, \eqref{f12}, \eqref{f13}, in the limit $t_\text{d1} \gg
t_\text{Lb},\:t_\text{Tb}$.  

The mean velocity $v_\text{run} = z_\text{run} / t_\text{run}$, where
$t_\text{run}$ is the average run time.  To calculate the latter, we
note that the mean waiting period (when both heads are bound
to actin) is $t_\text{d1} t_\text{d2}/(t_\text{d1}+t_\text{d2}) = g
t_\text{d1}/(1+g)$, while the mean binding times for the TH/LH are
$t_\text{Tb}$ [Eq.~\eqref{f10d}] and $t_\text{Lb}$ [Eq.~\eqref{f12d}]
respectively.  Then $t_\text{run}$ for $t_\text{d1} \gg
t_\text{Lb},\:t_\text{Tb}$ is given by:
\begin{equation}\label{f20}
t_\text{run} = \frac{{\cal P}_\text{f}+{\cal P}_\text{Ts}}{{\cal P}_\text{t}(1-{\cal P}_\text{t})} \left(\frac{g}{1+g}t_\text{d1}+t_\text{Tb}\right) + \frac{{\cal P}_\text{b}+{\cal P}_\text{Ls}}{{\cal P}_\text{t}(1-{\cal P}_\text{t})} \left(\frac{g}{1+g}t_\text{d1}+t_\text{Lb}\right) \approx \frac{g t_\text{d1}^2}{t_\text{Lb}+ g t_\text{Tb}},
\end{equation}
where the first term is the contribution of steps involving TH
detachment, and the second term those involving LH detachment. The
resulting expression for $v_\text{run}$ is:
\begin{equation}\label{f21}
v_\text{run} = \frac{z_\text{run}}{t_\text{run}} \approx \frac{\Delta}{t_\text{d1}}\left( \frac{1}{1+b\alpha} - \frac{\alpha}{g(b+\alpha)}\right).
\end{equation}
Eqs.~\eqref{f18}-\eqref{f21} are reproduced as Eq.~(9) in the main
text.

\section{Equilibrium probability of myosin end-point fluctuations}

\begin{figure}[t]
\centering\includegraphics*[width=0.7\textwidth]{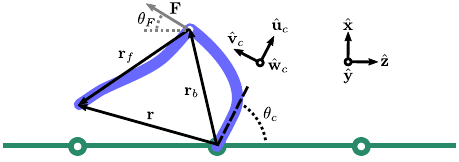}
\caption{Schematic diagram for the polymer model of MyoV, defining
  the free end-point vector $\mb{r}$ and the end-to-end vectors for
  the free ($\mb{r}_f$) and bound ($\mb{r}_b$) legs, respectively.
  The unit vector $\hat{\mb{u}}_c$ is the direction of the end-tangent
  constraint on the bound leg, and together with the two orthogonal
  unit vectors $\hat{\mb{v}}_c$ and $\hat{\mb{w}}_c$ it forms a set of
  axes tilted at an angle $\theta_c$ from the
  $(\hat{\mb{x}},\hat{\mb{y}},\hat{\mb{z}})$ axes, where $\hat{\mb{z}}$
  is oriented along the actin filament.}\label{diag}
\end{figure}

The equilibrium probability ${\cal P}(\mb{r})$ of finding the MyoV
free end at position $\mb{r}$ (Fig.~\ref{diag}), needed to calculate
$t^+_\text{fp}$ in Eq.~\eqref{f9}, can be obtained from calculating
the end-to-end vector probabilities of the bound leg, ${\cal
  P}_b(\mb{r}_b)$, and the free leg, ${\cal P}_f(\mb{r}_f)$.  Since
$\mb{r}$ is the sum of the end-to-end vectors of the legs, $\mb{r} =
\mb{r}_b + \mb{r}_f$, ${\cal P}(\mb{r})$ can be written as a
convolution of the two leg probabilities,
\begin{equation}\label{s1}
{\cal P}(\mb{r}) = \int d\mb{r}_b\int d\mb{r}_f\, {\cal P}_b(\mb{r}_b) {\cal P}_f(\mb{r}_f) \delta(\mb{r}-\mb{r}_b-\mb{r}_f).
\end{equation}
Each leg is a inextensible semiflexible polymer of contour length $L$
and persistence length $l_p$~\cite{Kratky49}, and one end of the bound
leg is fixed at the origin $\mb{r}=0$.  The bound leg has two
energetic contributions not present for the free leg: (i) the tangent
vector of the bound leg at the origin, $\hat{\mb{u}}_0$, is subject to
a harmonic constraint with energy ${\cal H}_c = \frac{1}{2}k_B T \nu_c
(\hat{\mb{u}}_0 - \hat{\mb{u}}_c)^2$, where $\nu_c$ and
$\hat{\mb{u}}_c$ are the strength and direction of the angle
constraint respectively ($\hat{\mb{v}}$ denotes a unit vector,
meaning $|\hat{\mb{v}}| = 1$); (ii) a load force $\mb{F}$ is applied at the
other end of the bound leg, where it joins the free leg.  The force is
oriented at an angle $\theta_F$ clockwise from the $-\hmb{z}$ axis, as
shown in Fig.~\ref{diag}.  The axis $\hat{\mb{z}}$ is oriented from
the $-$ to $+$ ends of the actin filament.  Both of these energetic
contributions will lead to an overall tension in the bound leg that
has to be accounted for in calculating the probability ${\cal
  P}_b(\mb{r}_b)$.  In the following subsections, we present
approximate analytical expressions for the leg probabilities ${\cal
  P}_f(\mb{r}_f)$ and ${\cal P}_b(\mb{r}_b)$, justifying them by
comparison with exact results for the first and second moments of the
equilibrium probabilities.  In the final subsection, we take the
individual leg results and use Eq.~\eqref{s1} to derive a complete
analytical expression for ${\cal P}(\mb{r})$, which is needed to
calculate the first passage times (Eq.~\eqref{f9}).

\vspace{1em}

\noindent {\bf Equilibrium end-to-end probability of the free leg.} We
start with the simpler case of the free leg, which is not under
tension.  There is no exact closed form analytical expression for the
end-to-end vector probability ${\cal P}_f(\mb{r}_f)$ of a semiflexible
polymer (though the moments of the probability distribution are known
analytically~\cite{Kratky49,Saito67}, as illustrated below).
Mean-field theory, however, provides an excellent approximation to the
distribution~\cite{Thirum98},
\begin{equation}\label{s2}
{\cal P}_f (\mb{r}_f) = A_f \xi_f^{-9/2} \exp\left(-\frac{3\kappa}{4 \xi_f}\right),
\end{equation}
where $\kappa = L/l_p$, $\xi_f = 1-r_f^2/L^2$, and $A_f$ is a
normalization constant.  The end-to-end vector $\mb{r}_f$ can be
specified by polar and azimuthal angles $\theta_f$ and $\phi_f$, and
the dimensionless radial variable $\xi_f$, which can only take on
values between 0 and 1 for an inextensible polymer, since $r_f \le
L$.  In this coordinate system the normalization
condition for the probability is:
\begin{equation}\label{s3}
1= \frac{L^3}{2}\int_0^1 d\xi_f (1-\xi_f)^{1/2} \int_0^\pi d\theta_f \int_0^{2\pi} d\phi_f\, {\cal P}_f(\mb{r}_f).
\end{equation}
The normalization constant $A_f$ is given by,
\begin{equation}\label{s4}
A_f = \frac{9 \sqrt{3} e^{3 \kappa/4} \kappa^{7/2}}{8 \pi ^{3/2} L^3 \left(3 \kappa^2+12 \kappa+20\right)}.
\end{equation}
In the stiff limit of large persistence length ($\kappa \to 0$), the
probability in Eq.~\eqref{s2} goes to a delta function at $r_f = L$,
as is appropriate for a rigid rod of length $L$.  In the opposite
limit of flexible chains ($\kappa \to \infty$), the probability goes
to a Gaussian centered at $\mb{r}=0$.  Throughout the entire range of
$\kappa$, the second moment of the probability distribution, $\langle
r_f^2 \rangle = 2L^2 (3 \kappa+10)/(3 \kappa^2+12 \kappa+20)$, is
within $1\%$ of the exact result $\langle r_f^2
\rangle_\text{exact} = 2 L^2\kappa^{-2} (\kappa -1 +
e^{-\kappa})$~\cite{Kratky49,Saito67}.  (The first moment $\langle
\mb{r}_f \rangle$ is trivially equal to zero in both the exact and
approximate cases because of the radial symmetry of the distribution.)
The approximation of Eq.~\eqref{s2} thus captures the physical
features of the stiff and flexible limits and is reasonably accurate
for our purposes.

\vspace{1em}

\noindent {\bf Equilibrium end-to-end probability of the bound leg at
  zero load.}  We first consider the bound leg in the absence of load
on the joint ($F=0$).  Our expression for ${\cal P}_b(\mb{r}_b)$
should reduce to the free leg probability of Eq.~\eqref{s2} in the
limit of zero constraint strength $\nu_c = 0$.  For $\nu_c \ne 0$, we
assume the effect of the end-tangent constraint can be approximated by
the following ansatz,
\begin{equation}\label{s5}
{\cal P}_b (\mb{r}_b) = A_b \xi_b^{-9/2} \exp\left(-\frac{3\kappa}{4 \xi_b} + {\cal T} \hat{\mb{u}}_c \cdot \hat{\mb{r}}_b\right),
\end{equation}
where $\xi_b = 1-r_b^2/L^2$, $A_b$ is a normalization constant, and
${\cal T}$ is an unknown function of $\nu_c$ to be determined later,
satisfying ${\cal T} = 0$ at $\nu_c =0$.  Eq.~\eqref{s5} is identical in form to
Eq.~\eqref{s2}, except for the additional ${\cal T}$ term in the
exponential, which acts as an effective tension along $\hat{\mb{u}}_c$
due to the end-tangent constraint.  The normalization constant $A_b$ is given by:
\begin{equation}\label{s6}
A_b = A_f \frac{{\cal T}}{\sinh {\cal T}}.
\end{equation}
We choose ${\cal T}$ so that the first and second
moments of the probability distribution of Eq.~\eqref{s5} closely
agree with the exact values for a semiflexible polymer under a
harmonic end-tangent constraint.  Because the analytical expressions
for these exact values are not available in the literature, we derive
them in the following way.  We start by noting that the bound leg
end-to-end vector $\mb{r}_b = \int_0^L ds\, \hat{\mb{u}}(s)$, where
$\hat{\mb{u}}(s) = d\mb{r}(s)/ds$ is the tangent vector at position
$s$ along the bound leg chain contour $\mb{r}(s)$, $0\le s \le L$.
The tangent vectors for an inextensible chain all have unit length.
The equilibrium statistics of $\hat{\mb{u}}(s)$ for a semiflexible
polymer are governed by the Green's function $G(\hat{\mb{u}},
\hat{\mb{u}}^\prime;s-s^\prime)$, which describes the probability that
a chain with tangent vector $\hat{\mb{u}}(s) = \hat{\mb{u}}$ will have
tangent vector $\hat{\mb{u}}(s^\prime) = \mb{u}^\prime$ at some
position $s^\prime \ge s$.  This Green's function has an exact
spherical harmonic expansion of the form~\cite{Saito67},
\begin{equation}\label{s7}
G(\hat{\mb{u}},\hat{\mb{u}}^\prime;s-s^\prime) = \sum_{l,m} e^{-\frac{l(l+1)}{2 l_p}(s^\prime-s)} Y^\ast_{lm}(\hat{\mb{u}})Y_{lm}(\hat{\mb{u}}^\prime).
\end{equation}
For the initial tangent vector $\hat{\mb{u}}_0 \equiv \hat{\mb{u}}(0)$
at $s=0$, where the bound leg is attached to the actin, the harmonic
constraint leads to a probability distribution ${\cal
  P}_c(\hat{\mb{u}}_0)$ given by:
\begin{equation}\label{s8}
\begin{split}
{\cal P}_c(\hat{\mb{u}}_0) &= \frac{\nu_c}{2\pi (1-e^{-2\nu_c})} \exp\left(-\frac{\nu_c}{2}(\hat{\mb{u}}_0 - \hat{\mb{u}}_c)^2 \right)\\
&= \sqrt\frac{\pi \nu_c}{2}\frac{1}{\sinh \nu_c} \sum_{l,m} I_{l+1/2}(\nu_c)Y^\ast_{lm}(\hat{\mb{u}}_c)Y_{lm}(\hat{\mb{u}}_0).
\end{split}
\end{equation}
In the first line the prefactor in front of the exponential is a
normalization constant.  In the second line, we have rewritten the
exponential in a spherical harmonic expansion~\cite{Chiu94} involving
modified spherical Bessel functions of the first kind $I_\nu(x)$.  This form will
facilitate carrying out the moment integrals below.

Let $\hat{\mb{t}}$ be one of the three orthogonal unit vectors
$\hat{\mb{u}}_c$, $\hat{\mb{v}}_c$, or $\hat{\mb{w}}_c$, defined in
Fig.~\ref{diag}.  These axes, with $\hat{\mb{u}}_c$ being the
constraint direction, are the easiest to work with for moment
calculations.  Using the definitions of
$G(\hat{\mb{u}},\hat{\mb{u}}^\prime;s-s^\prime)$ and ${\cal
  P}_c(\hat{\mb{u}}_0)$ above, the first and second order moments with
respect to one of the axes $\hat{\mb{t}}$ can be written as,
\begin{equation}\label{s9}
\begin{split}
\langle \hat{\mb{t}} \cdot \mb{r}_b \rangle_\text{exact} &= \left\langle 
\int_0^L ds\,\hat{\mb{t}}\cdot\hat{\mb{u}}(s)\right\rangle_\text{exact}\\
&=  \int_0^L ds \int
d\hat{\mb{u}}_0 \int d\hat{\mb{u}}\,
{\cal P}_c(\hat{\mb{u}}_0) G(\hat{\mb{u}}_0,\hat{\mb{u}};s)\, \hat{\mb{t}}\cdot \hat{\mb{u}},\\
\langle (\hat{\mb{t}} \cdot \mb{r}_b)^2 \rangle_\text{exact} &= \left\langle 
\int_0^L ds\int_0^L ds^\prime \,\hat{\mb{t}}\cdot\hat{\mb{u}}(s)\,\hat{\mb{t}}\cdot\hat{\mb{u}}(s^\prime)\right\rangle_\text{exact}\\
&= 2 \int_0^L ds  \int_s^L ds^\prime \int
d\hat{\mb{u}}_0 \int d\hat{\mb{u}}\int d\hat{\mb{u}}^\prime \\
&\qquad\qquad{\cal P}_c(\hat{\mb{u}}_0) G(\hat{\mb{u}}_0,\hat{\mb{u}};s)\, \hat{\mb{t}}\cdot \hat{\mb{u}}\, G(\hat{\mb{u}},\hat{\mb{u}}^\prime;s^\prime-s)\, \hat{\mb{t}}\cdot \hat{\mb{u}}^\prime.\\
\end{split}
\end{equation}
By using Eqs.~\eqref{s7}-\eqref{s8} and the properties of spherical
harmonics, the integrals in Eq.~\eqref{s9} can be carried out exactly
to yield the moments for any axis $\hat{\mb{t}}$.  Let us define the
average end-to-end component parallel to the constraint direction,
$\mu^\text{exact}_\parallel \equiv \langle \hmb{u}_c \cdot \mb{r}_b
\rangle_\text{exact}$ (the first moments along $\hmb{v}_c$ and
$\hmb{w}_c$ are zero).  Similarly, define the parallel and
perpendicular end-to-end standard deviations, $\sigma^\text{exact}_\parallel \equiv
(\langle (\hmb{u}_c \cdot \mb{r}_b)^2 \rangle_\text{exact} - \langle \hmb{u}_c \cdot
\mb{r}_b \rangle_\text{exact}^2)^{1/2}$, $\sigma^\text{exact}_\perp \equiv \langle (\hmb{v}_c
\cdot \mb{r}_b)^2 \rangle_\text{exact}^{1/2} = \langle (\hmb{w}_c \cdot \mb{r}_b)^2 \rangle_\text{exact}^{1/2}$.
The results for these three quantities are:
\begin{equation}\label{s10}
\begin{split}
\mu_\parallel^\text{exact} &= L\kappa^{-1}\left(1-k \right) {\cal L}\left(\nu _c\right),\\
\sigma_\parallel^\text{exact}&=\frac{L\kappa^{-1}}{3}\left(2 \left(3 \kappa +k^3-1\right)-9 (k-1)^2 {\cal L}^2\left(\nu _c\right)-\frac{6 (k+2) (k-1)^2 {\cal L}\left(\nu _c\right)}{\nu _c}\right)^{1/2},\\
\sigma_\perp^\text{exact} &= \frac{L\kappa^{-1}}{3} \left(6 \kappa -k^3+9 k-8+\frac{3 \left(k^3-3 k+2\right) {\cal L}\left(\nu _c\right)}{\nu _c}\right)^{1/2},
\end{split}
\end{equation}
where $k \equiv \exp(-\kappa)$ and ${\cal L}(\nu_c) \equiv \coth\nu_c - \nu_c^{-1}$ is the Langevin function.

\begin{figure}[t]
\centering\includegraphics*[width=0.95\textwidth]{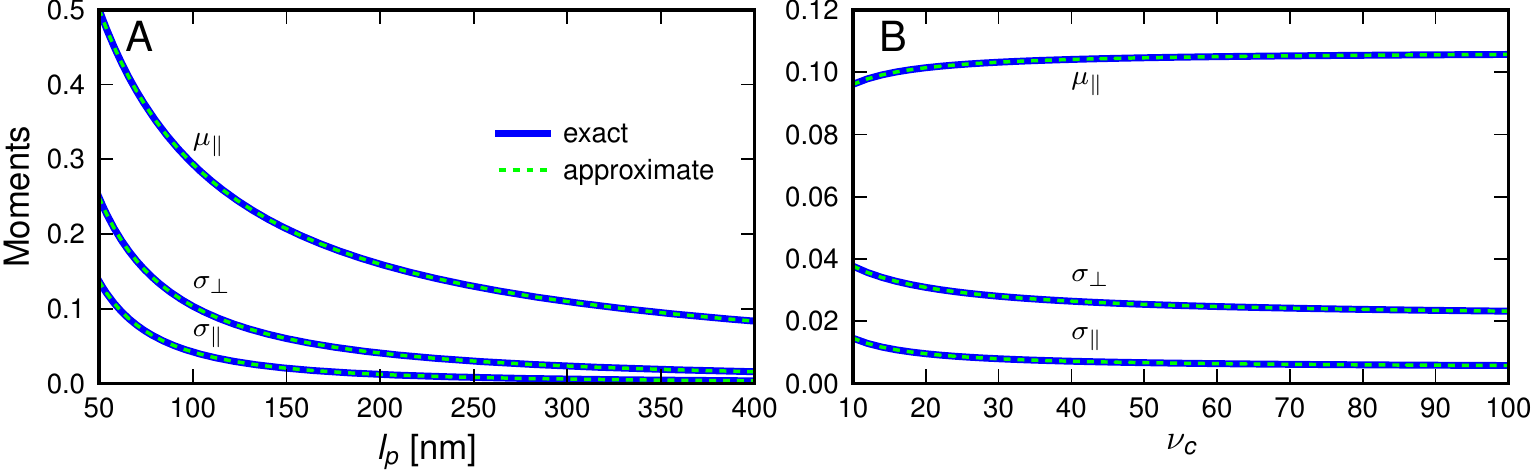}
\caption{First and second moments of the end-to-end vector
  distribution for the bound leg when $\mb{F}=0$, measured in units of
  leg persistence length $l_p$.  The exact values (solid lines) are
  given by Eq.~\eqref{s10}, while the approximate values (dashed
  lines) are taken from Eq.~\eqref{s11} with ${\cal T}$ defined by
  Eq.~\eqref{s13}.  (a) Moments as a function of persistence length
  $l_p$ for fixed constraint strength $\nu_c = 180$.  (b) Moments as a
  function of $\nu_c$ for fixed $l_p = 310$ nm.}\label{blavg}
\end{figure}

The corresponding moments calculated from the probability distribution in Eq.~\eqref{s5} are:
\begin{equation}\label{s11}
\begin{split}
\mu_\parallel &= \frac{L {\cal L}\left({\cal T}\right)}{\sqrt{\pi } \left(\frac{9}{4} \kappa  (\kappa +4)+15\right)}\left(\frac{3 \sqrt{\pi } (10-3 \kappa )}{2 k^{3/4}}\text{erfc}\frac{\sqrt{3 \kappa }}{2} +3 \sqrt{3 \kappa }
   (\kappa +5)\right),\\
\sigma_\parallel &= L\kappa^{-1}\left(\frac{2 \kappa ^2 (3 \kappa +10) \left({\cal T}-2 {\cal L}\left({\cal T}\right)\right)}{(3 \kappa  (\kappa +4)+20) {\cal T}} - \frac{\mu_\parallel^2}{L^2}\right)^{1/2},\\
\sigma_\perp &= L\kappa^{-1}\left(\frac{2 \kappa ^2 (3 \kappa +10) {\cal L}\left({\cal T}\right)}{(3 \kappa  (\kappa +4)+20) {\cal T}}\right)^{1/2}.
\end{split}
\end{equation}
To determine ${\cal T}$, we will set $\mu_\parallel$ from
Eq.~\eqref{s11} equal to $\mu_\parallel^\text{exact}$ from
Eq.~\eqref{s10}.  The resulting expression for ${\cal T}$ is:
\begin{equation}\label{s12}
{\cal T} = {\cal L}^{-1}\left(\frac{\sqrt{\pi } (3 \kappa  (\kappa +4)+20) (1-k) k^{3/4} {\cal L}\left(\nu _c\right)}{2 \kappa  \left(\sqrt{\pi } (10-3 \kappa )
   \text{erfc}\left(\frac{\sqrt{3\kappa }}{2}\right)+2 \sqrt{3\kappa } (\kappa +5) k^{3/4}\right)} \right).
\end{equation}
Since the inverse Langevin function ${\cal L}^{-1}(x)$ cannot be
expressed analytically, for the purposes of evaluation we use the
Pad\'e approximation ${\cal L}^{-1}(x) \approx
x(3-x^2)/(1-x^2)$~\cite{Cohen91}.  For the parameter regime $\kappa
\ll 1$ (large stiffness) and $\nu_c \gg 1$ (strong end-tangent
constraint), relevant to MyoV dynamics, Eq.~\eqref{s12} can be
further simplified to yield:
\begin{equation}\label{s13}
{\cal T} \approx 1+ \frac{20 \nu_c}{20+7\kappa \nu_c }.
\end{equation}
Eqs.~\eqref{s5}, \eqref{s6}, and \eqref{s13} completely describe the
end-to-end vector probability distribution for the bound leg at zero
load.  By construction the ${\cal T}$ of Eq.~\eqref{s13} leads to a
$\mu_\parallel$ that closely agrees with the exact value
$\mu_\parallel^\text{exact}$ from Eq.~\eqref{s10}.  In addition, the other
moments are also well reproduced by the approximate probability
distribution, as shown in Fig.~\ref{blavg}.  The exact and approximate
values differ by no more than 7\% over the entire parameter range of
$l_p$ and $\nu_c$ shown in the figure.  This range covers the most
likely parameters for MyoV dynamics, as discussed in the main
text.

\vspace{1em}

\noindent {\bf Equilibrium end-to-end probability of the bound leg
  under load.} In the presence of a load force $\mb{F}$, the
probability distribution in Eq.~\eqref{s5} is multiplied by a factor
of $\exp(\beta F r_b \hmb{F}\cdot \hmb{r}_b) = \exp(\beta F L
(1-\xi_b)^{1/2} \hmb{F}\cdot \hmb{r}_b)$.  In the stiff limit $\kappa
\ll 1$, the main contributions to the end-to-end vector probability
are for $\xi_b \ll 1$, since $r_b$ approaches $L$, the leg contour
length.  Thus, the contribution of the load can be approximated as
$\exp(\beta F L \hmb{F}\cdot \hmb{r}_b)$.  With this approximation,
the overall form of Eq.~\eqref{s5} and \eqref{s6} is preserved under
load, with the substitutions ${\cal T} \to {\cal T}^\prime$ and
$\hmb{u}_c \to \hmb{u}^\prime_c$.  The probability distribution
becomes:
\begin{equation}\label{s5f}
{\cal P}_b (\mb{r}_b) = A_b \xi_b^{-9/2} \exp\left(-\frac{3\kappa}{4 \xi_b} + {\cal T}^\prime \hat{\mb{u}}^\prime_c \cdot \hat{\mb{r}}_b\right),
\end{equation}
\begin{equation}\label{s6f}
A_b = A_f \frac{{\cal T}^\prime}{\sinh {\cal T}^\prime},
\end{equation}
where the new effective tension along the leg, written in terms of its
$\hmb{x}$ and $\hmb{z}$ components, is:
\begin{equation}\label{s14}
{\cal T}^\prime = \sqrt{({\cal T}^\prime_x)^2+({\cal T}^\prime_z)^2},
\quad {\cal T}^\prime_x = {\cal T} \sin\theta_c + \beta F L
\sin\theta_F, \quad {\cal T}^\prime_z = {\cal T} \cos\theta_c - \beta
F L \cos\theta_F.
\end{equation}
The new effective tension direction is $\hmb{u}^\prime_c =
\sin\theta_c^\prime\hmb{x} + \cos\theta_c^\prime \hmb{z}$, which is
oriented at an angle $\theta_c^\prime$ from the $\hmb{z}$ axis,
\begin{equation}\label{s15}
\theta_c^\prime = \theta_c + \tan^{-1}\left(\frac{\beta FL \sin (\theta_c+\theta_F)}{{\cal T}-\beta F L \cos (\theta_c+\theta_F)} \right).
\end{equation}
\vspace{1em}

\noindent {\bf Combining the individual leg probabilities to find the
  total end-to-end vector probability distribution.}  The final step
in the derivation of ${\cal P}(\mb{r})$ is to evaluate Eq.~\eqref{s1}.
Using ${\cal P}_f$ from Eq.~\eqref{s2} and ${\cal P}_b$ from
Eq.~\eqref{s5f}, the convolution integral in Eq.~\eqref{s1} has the
form:
\begin{equation}\label{s16}
\begin{split}
{\cal P}(\mb{r}) &= A_f A_b \int d\mb{r}_b\int d\mb{r}_f\, \xi_f^{-9/2} \xi_b^{-9/2} \exp\left(-\frac{3\kappa}{4 \xi_f} -\frac{3\kappa}{4 \xi_b} + {\cal T}^\prime \hat{\mb{u}}^\prime_c \cdot \hat{\mb{r}}_b\right)  \delta(\mb{r}-\mb{r}_b-\mb{r}_f)\\
&= A_f A_b \int d\mb{r}_b\, \xi^{-9/2}_f \xi_b^{-9/2} \exp\left(-\frac{3\kappa}{4 \xi_f} -\frac{3\kappa}{4 \xi_b} + {\cal T}^\prime \hat{\mb{u}}^\prime_c \cdot \hat{\mb{r}}_b\right).
\end{split}
\end{equation}
In the second step we have carried out the integration over the free
leg end-to-end vector $\mb{r}_f$, with the delta function making the
radial variable $\xi_f = 1-r_f^2/L^2$ a function of $\mb{r}$ and $\mb{r}_b$,
\begin{equation}\label{s17}
\xi_f = 1- \frac{r^2+r_b^2-2r r_b\cos \theta_b}{L^2},
\end{equation}
where $\theta_b$ is the angle between $\mb{r}$ and $\mb{r}_b$.  Since
we are interested in probabilities of finding the free end of MyoV along the actin filament, let us confine the rest of the
calculation to $\mb{r} = z \hmb{z}$, where $-2L \le z \le 2L$ (since
this is the maximum range which the two-legged structure of total
contour length $2L$ can access).  The unit vector $\hmb{r}_b$ can be
represented in spherical coordinates by the polar and azimuthal angles
$(\theta_b, \phi_b)$, and $\hmb{u}_c^\prime$ by
$(\theta_c^\prime,\phi_c^\prime=0)$.  Thus:
\begin{equation}\label{s18}
\hmb{u}_c^\prime \cdot \hmb{r}_b = \cos \theta_b \cos\theta_c^\prime + \cos\phi_b \sin \theta_b \sin \theta_c^\prime.
\end{equation}
Writing the integration element in Eq.~\eqref{s16} as $d\mb{r}_b =
r_b^2 d\cos \theta_b d\phi_b$, we can carry out the integral over $\phi_b$ using Eq.~\eqref{s18}.  The result is:
\begin{equation}\label{s19a}
\begin{split}
{\cal P}(z \hmb{z}) &= 2\pi A_f A_b \int_0^L r_b^2 dr_b \int_{-1}^1 d\cos \theta_b\, \xi^{-9/2}_f \xi_b^{-9/2} \exp\left(-\frac{3\kappa}{4\xi_f} -\frac{3\kappa}{4 \xi_b} + {\cal T}^\prime_z \cos\theta_b\right)\\
&\qquad\qquad\qquad\qquad\cdot I_0\left({\cal T}^\prime_x\sin\theta_b\right),
\end{split}
\end{equation}
where $I_0(x)$ is the zeroth-order modified Bessel function of the
first kind.  To simplify the integration, we will change variables
from $(r_b, \cos\theta_b)$ to $(\xi_b,\xi_f)$.  From the definitions
of $\xi_b$, $\xi_f$, and Eq.~\eqref{s17}, the two sets of variables
are related by:
\begin{equation}\label{s20a}
r_b = L\sqrt{1-\xi_b}, \quad \cos\theta_b = \frac{z^2 + L^2(\xi_f-\xi_b)}{2 z L \sqrt{1-\xi_b}},
\end{equation}
leading to a Jacobian determinant $|\det J| = L^2/(4|z|(1-\xi_b))$ for
the change of variables.  Using these relations, Eq.~\eqref{s19a}
becomes:
\begin{equation}\label{s19}
\begin{split}
{\cal P}(z \hmb{z}) &= \frac{L^4\pi A_f A_b}{2|z|} \int_0^{u_b(z)} d\xi_b \int_0^{u_f(z,u_b)} d\xi_f\, \xi^{-9/2}_f \xi_b^{-9/2}\\
&\qquad\qquad\qquad\qquad\cdot \exp\left(-\frac{3\kappa}{4\xi_f} -\frac{3\kappa}{4 \xi_b} + {\cal T}^\prime_z \frac{z^2+L^2(\xi_f-\xi_b)}{2 zL\sqrt{1-\xi_b}} \right)\\
&\qquad\qquad\qquad\qquad\cdot I_0\left({\cal T}^\prime_x
\sqrt{1-\left(\frac{z^2+L^2(\xi_f-\xi_b)}{2 zL\sqrt{1-\xi_b}}\right)^2}
\right),
\end{split}
\end{equation}
where the upper limits of integration are given by:
\begin{equation}\label{s20}
u_b(z) = \frac{2|z|}{L} - \frac{z^2}{L^2}, \quad u_f(z,\xi_b) = \xi_b  +\frac{2|z|\sqrt{1-\xi_b}}{L}- \frac{z^2}{L^2}.
\end{equation}
In the stiff limit $\kappa \to 0$, the main contributions to the
integral come from $\xi_b \ll 1$ and $\xi_f \ll 1$.  Additionally, the
location of the binding sites we consider, $|z| = 36$ nm, are
comparable to the leg contour length $L = 35$ nm.  We can then
approximately carry out the integral in Eq.~\eqref{s19} by replacing
the integration limits $u_b \to 1$, $u_f \to 1$, and substituting
\begin{equation}\label{s21}
\frac{z^2+L^2(\xi_f-\xi_b)}{2 zL\sqrt{1-\xi_b}} \to \frac{z}{2 L}.
\end{equation}
With these approximations, the integral in Eq.~\eqref{s19} evaluates
to:
\begin{equation}\label{s22}
\begin{split}
{\cal P}(z \hmb{z}) &\approx
\frac{8 \pi  L^4 A_f A_b}{729 \kappa^7 |z|} \left(20 \sqrt{3 \pi } e^{3 \kappa /4} \text{erfc}\left(\frac{\sqrt{3\kappa }}{2}\right)+3 \sqrt{\kappa } (\kappa  (3 \kappa
   +10)+20)\right)^2\\
&\qquad \cdot I_0\left({\cal T}^\prime_x\sqrt{1-\frac{z^2}{4 L^2}}  \right) e^{\frac{1}{2} \left(\frac{{\cal T}^\prime_z z}{L}-3 \kappa \right)}.
\end{split}
\end{equation}
Upon substituting in Eq.~\eqref{s4} for $A_f$, Eq.~\eqref{s6} for $A_b$, and
expanding ${\cal P}(z \hmb{z})$ up to second order in $\kappa$, we get
the final, simplified form of the probability.  The result evaluated
at $z=\pm \Delta$ is given by Eq.~(20) of the main text:
\begin{equation}\label{s23}
\begin{split}
{\cal P}(\mb{r}_\pm) &\approx \frac{(3 \kappa  (7 \kappa +20)+200) {\cal T}^\prime}{1600 \pi  L^2 \Delta \sinh {\cal T}^\prime} I_0\left({\cal T}^\prime_x \sqrt{1- \frac{\Delta^2}{4 L^2}}  \right)   e^{\pm \frac{{\cal T}^\prime_z \Delta}{2 L}}.
\end{split}
\end{equation}
Together with Eq.~\eqref{s13} for ${\cal T}$ and Eq.~\eqref{s14} for
${\cal T}^\prime$, we now have a complete analytical expression for
the probability distribution of the MyoV free end at any location
along the actin filament axis.

\begin{figure}
\centering\includegraphics*[width=0.95\textwidth]{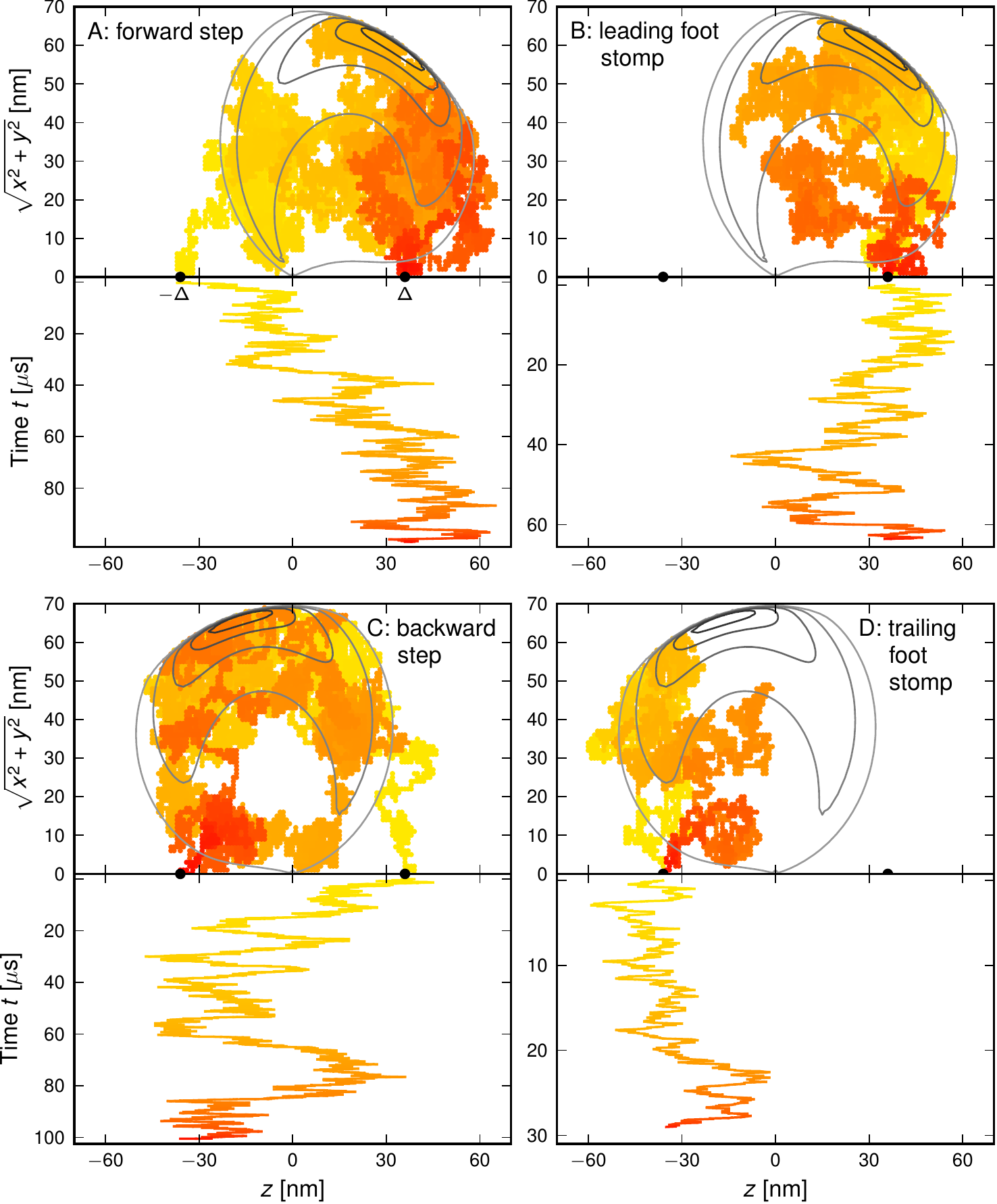}
  \caption{ Sample trajectories of the end-to-end vector
    $\mb{r}=(x,y,z)$ for each of the four MyoV kinetic pathways,
    calculated from a numerical solution~\cite{Wang03} to the
    Fokker-Planck equation with diffusivity $D_\text{h}$ and an energy
    landscape $U(\mb{r}) = -k_BT \log {\cal P}(\mb{r})$, with ${\cal
      P}(\mb{r})$ given by Eq.~\eqref{s16}.  The top panels for each
    pathway show the trajectories in terms of $z$ (the distance along
    actin) vs. $\sqrt{x^2+y^2}$, with colors from yellow to red
    denoting progress in time.  The bottom panels show the
    corresponding $z(t)$ for the trajectory, using the same color
    coding.  Superimposed on the top panels are contour lines of
    ${\cal P}(\mb{r})$ for probabilities $1,2,\ldots,5 \cdot 10^{-4}$
    nm$^{-3}$ (light gray to dark gray).  The pathways in A and B are
    at $F=0$ pN, while those for C and D are at $F=2$ pN, and hence
    the ${\cal P}(\mb{r})$ distribution in the latter cases is shifted
    in the $-\hmb{z}$ direction.}\label{traj}
\end{figure}

An analogous approach can be used to find ${\cal
    P}(\mb{r})$ analytically at any $\mb{r}$, not just along
  $\hmb{z}$.  The resulting 3D probability distribution allows us to
  generate sample diffusive trajectories for the end-to-end vector
  $\mb{r}$ in various MyoV kinetic pathways, as shown in
  Fig.~\ref{traj}.  These are numerical solutions to the Fokker Planck
  equation~\cite{Wang03} for diffusion along an energy surface
  $U(\mb{r}) = -k_B T \log {\cal P}(\mb{r})$ with diffusivity
  $D_\text{h}$.

\section{Stall force}

Based on the earlier results for the step probabilities and first
passage times, one can derive a simple expression for the stall force
$F_\text{stall}$, defined by the condition that backward and forward
step probabilities are equal, ${\cal P}_\text{f} = {\cal
  P}_\text{b}$ at $F = F_\text{stall}$.  From Eqs.~\eqref{f10}
and \eqref{f13}, the ratio of the two probabilities is:
\begin{equation}\label{st1}
\begin{split}
\frac{{\cal P}_\text{b}}{{\cal P}_\text{f}} &= \frac{\alpha (1+b\alpha)(t_\text{d1}+t_\text{h})(t_\text{d1}+t_\text{Tb}-t_\text{h})}{g(b+\alpha)t_\text{d1} (t_\text{d1}+t_\text{Lb})}\\
&\approx g^{-1} \frac{\alpha (1+ b \alpha)}{b+\alpha},
\end{split}
\end{equation}
The approximation in the second line is valid when $t_\text{d1} \gg
t_\text{Tb},\: t_\text{Lb}$, which is typically the case.

Setting the right-hand side of Eq.~\eqref{st1} equal to 1, we can
solve for the value $\alpha = \alpha_\text{stall}$ at the stall force,
\begin{equation}\label{st2}
\alpha_\text{stall} = \frac{g-1+\sqrt{(g-1)^2 +4gb^2}}{2 b}.
\end{equation}
Using Eq.~\eqref{f9} for $t^\pm_\text{fp}$, Eq.~\eqref{s23} for the
equilibrium free end probability ${\cal P}(\mb{r}_\pm)$, and the
definition of ${\cal T}^\prime$ from Eq.~\eqref{s14}, we can rewrite
Eq.~\eqref{st2} as follows:
\begin{equation}\label{st3}
\begin{split}
\frac{g-1+\sqrt{(g-1)^2 +4gb^2}}{2 b} = \alpha_\text{stall} &= \left.\frac{{\cal P}(\mb{r}_-)}{{\cal P}(\mb{r}_+)}\right|_{F=F_\text{stall}}= \exp\left(-\frac{\Delta {\cal T}}{L} \cos\theta_c + \beta \Delta F_\text{stall} \cos\theta_F\right).
\end{split}
\end{equation}
This equation can be directly solved for $F_\text{stall}$, giving Eq.~(10) of the main text,
\begin{equation}\label{st4}
F_\text{stall} = \frac{k_B T}{\cos \theta_F}\left(\frac{{\cal T}}{L} \cos\theta_c + \frac{1}{\Delta} \log \frac{g-1+\sqrt{(g-1)^2 +4gb^2}}{2 b}\right).
\end{equation}

\end{widetext}

\begin{thebibliography}{10}

\bibitem{Sivaramakrishnan07}
Spudich JA, Sivaramakrishnan S
\newblock (2010) Myosin {VI}: an innovative motor that challenged the swinging
  lever arm hypothesis.
\newblock \emph{Nat. Rev. Mol. Cell. Biol.} 11:128--137.

\bibitem{Reck-Peterson00}
Reck-Peterson SL, Provance DW, Mooseker MS, Mercer JA
\newblock (2000) Class {V} myosins.
\newblock \emph{Biochim. Biophys. Acta-Mol. Cell Res.} 1496:36--51.

\bibitem{Shiroguchi11}
Shiroguchi K, {et~al.}
\newblock (2011) Direct {Observation} of the {Myosin} {Va} {Recovery} {Stroke}
  {That} {Contributes} to {Unidirectional} {Stepping} along actin.
\newblock \emph{PLoS. Biol.} 9:e1001031.

\bibitem{Mehta99}
Mehta AD, {et~al.}
\newblock (1999) Myosin-{V} is a processive actin-based motor.
\newblock \emph{Nature} 400:590--593.

\bibitem{Rief00}
Rief M, {et~al.}
\newblock (2000) Myosin-{V} stepping kinetics: {A} molecular model for
  processivity.
\newblock \emph{Proc. Natl. Acad. Sci. U.S.A.} 97:9482--9486.

\bibitem{Sakamoto00}
Sakamoto T, Amitani I, Yokota E, Ando T
\newblock (2000) Direct observation of processive movement by individual myosin
  {V} molecules.
\newblock \emph{Biochem. Biophys. Res. Commun.} 272:586--590.

\bibitem{Yildiz03}
Yildiz A, {et~al.}
\newblock (2003) Myosin {V} walks hand-over-hand: {Single} fluorophore imaging
  with 1.5-nm localization.
\newblock \emph{Science} 300:2061--2065.

\bibitem{Forkey03}
Forkey JN, Quinlan ME, Shaw MA, Corrie JET, Goldman YE
\newblock (2003) Three-dimensional structural dynamics of myosin {V} by
  single-molecule fluorescence polarization.
\newblock \emph{Nature} 422:399--404.

\bibitem{Sakamoto08}
Sakamoto T, Webb MR, Forgacs E, White HD, Sellers JR
\newblock (2008) Direct observation of the mechanochemical coupling in myosin
  {Va} during processive movement.
\newblock \emph{Nature} 455:128--U99.

\bibitem{Baker04}
Baker JE, {et~al.}
\newblock (2004) Myosin {V} processivity: Multiple kinetic pathways for
  head-to-head coordination.
\newblock \emph{Proc. Natl. Acad. Sci. U.S.A.} 101:5542--5546.

\bibitem{Pierobon09}
Pierobon P, {et~al.}
\newblock (2009) Velocity, {processivity,} and {individual} {steps} of {single}
  {myosin} {V} {molecules} in {live} cells.
\newblock \emph{Biophys. J.} 96:4268--4275.

\bibitem{Veigel02}
Veigel C, Wang F, Bartoo ML, Sellers JR, Molloy JE
\newblock (2002) The gated gait of the processive molecular motor, myosin {V}.
\newblock \emph{Nat. Cell Biol.} 4:59--65.

\bibitem{Rosenfeld04}
Rosenfeld SS, Sweeney HL
\newblock (2004) A model of myosin {V} processivity.
\newblock \emph{J. Biol. Chem.} 279:40100--40111.

\bibitem{Veigel05}
Veigel C, Schmitz S, Wang F, Sellers JR
\newblock (2005) Load-dependent kinetics of {myosin-V} can explain its high
  processivity.
\newblock \emph{Nat. Cell Biol.} 7:861--869.

\bibitem{Purcell05}
Purcell TJ, Sweeney HL, Spudich JA
\newblock (2005) A force-dependent state controls the coordination of
  processive myosin {V}.
\newblock \emph{Proc. Natl. Acad. Sci. U.S.A.} 102:13873--13878.

\bibitem{Kad08}
Kad NM, Trybus KM, Warshaw DM
\newblock (2008) Load and {P(i)} control flux through the branched kinetic
  cycle of myosin {V}.
\newblock \emph{J. Biol. Chem.} 283:17477--17484.

\bibitem{Uemura04}
Uemura S, Higuchi H, Olivares AO, {De La Cruz} EM, Ishiwata S
\newblock (2004) Mechanochemical coupling of two substeps in a single myosin
  {V} motor.
\newblock \emph{Nat. Struct. Mol. Biol.} 11:877--883.

\bibitem{Gebhardt06}
Gebhardt JCM, Clemen AEM, Jaud J, Rief M
\newblock (2006) Myosin-{V} is a mechanical ratchet.
\newblock \emph{Proc. Natl. Acad. Sci. U.S.A.} 103:8680--8685.

\bibitem{Cappello07}
Cappello G, {et~al.}
\newblock (2007) Myosin {V} stepping mechanism.
\newblock \emph{Proc. Natl. Acad. Sci. U.S.A.} 104:15328--15333.

\bibitem{Kodera10}
Kodera N, Yamamoto D, Ishikawa R, Ando T
\newblock (2010) Video imaging of walking myosin {V} by high-speed atomic force
  microscopy.
\newblock \emph{Nature} 468:72--77.

\bibitem{Syed06}
Syed S, Snyder GE, Franzini-Armstrong C, Selvin PR, Goldman YE
\newblock (2006) Adaptability of myosin {V} studied by simultaneous detection
  of position and orientation.
\newblock \emph{Embo J.} 25:1795--1803.

\bibitem{Beausang13}
Beausang JF, Shroder DY, Nelson PC, Goldman YE
\newblock (2013) Tilting and wobble of myosin {V} by high-speed single-molecule
  polarized fluorescence microscopy.
\newblock \emph{Biophys. J.} 104:1263--1273.

\bibitem{Dunn07}
Dunn AR, Spudich JA
\newblock (2007) Dynamics of the unbound head during myosin {V} processive
  translocation.
\newblock \emph{Nat. Struct. Mol. Biol.} 14:246--248.

\bibitem{Shiroguchi07}
Shiroguchi K, Kinosita K
\newblock (2007) Myosin {V} walks by lever action and {Brownian} motion.
\newblock \emph{Science} 316:1208--1212.

\bibitem{Komori07}
Komori Y, Iwane AH, Yanagida T
\newblock (2007) Myosin-{V} makes two brownian 90 degrees rotations per 36-nm
  step.
\newblock \emph{Nat. Struct. Mol. Biol.} 14:968--973.

\bibitem{Howard96}
Howard J, Spudich JA
\newblock (1996) Is the lever arm of myosin a molecular elastic element?
\newblock \emph{Proc. Natl. Acad. Sci. U.S.A.} 93:4462--4464.

\bibitem{Moore04}
Moore JR, Krementsova EB, Trybus KM, Warshaw DM
\newblock (2004) Does the myosin {V} neck region act as a lever?
\newblock \emph{J. Muscle Res. Cell Motil.} 25:29--35.

\bibitem{Vilfan05}
Vilfan A
\newblock (2005) Elastic lever-arm model for myosin {V}.
\newblock \emph{Biophys. J.} 88:3792--3805.

\bibitem{Kolomeisky03}
Kolomeisky AB, Fisher ME
\newblock (2003) A simple kinetic model describes the processivity of
  myosin-{V}.
\newblock \emph{Biophys. J.} 84:1642--1650.

\bibitem{Lan05}
Lan GH, Sun SX
\newblock (2005) Dynamics of {myosin-V} processivity.
\newblock \emph{Biophys. J.} 88:999--1008.

\bibitem{Skau06}
Skau KI, Hoyle RB, Turner MS
\newblock (2006) A kinetic model describing the processivity of myosin-{V}.
\newblock \emph{Biophys. J.} 91:2475--2489.

\bibitem{Tsygankov07}
Tsygankov D, Fisher ME
\newblock (2007) Mechanoenzymes under superstall and large assisting loads
  reveal structural features.
\newblock \emph{Proc. Natl. Acad. Sci. U.S.A.} 104:19321--19326.

\bibitem{Xu09}
Xu YZ, Wang ZS
\newblock (2009) Comprehensive physical mechanism of two-headed biomotor myosin
  {V}.
\newblock \emph{J. Chem. Phys.} 131:245104.

\bibitem{Bierbaum11}
Bierbaum V, Lipowsky R
\newblock (2011) Chemomechanical {coupling} and {motor} {cycles} of {myosin}
  {V}.
\newblock \emph{Biophys. J.} 100:1747--1755.

\bibitem{Craig09}
Craig EM, Linke H
\newblock (2009) Mechanochemical model for myosin {V}.
\newblock \emph{Proc. Natl. Acad. Sci. U.S.A.} 106:18261--18266.

\bibitem{Clemen05}
Clemen AEM, {et~al.}
\newblock (2005) Force-dependent stepping kinetics of myosin-{V}.
\newblock \emph{Biophys. J.} 88:4402--4410.

\bibitem{Walker00}
Walker ML, {et~al.}
\newblock (2000) Two-headed binding of a processive myosin to {F}-actin.
\newblock \emph{Nature} 405:804--807.

\bibitem{DeLaCruz99}
{De La Cruz} EM, Wells AL, Rosenfeld SS, Ostap EM, Sweeney HL
\newblock (1999) The kinetic mechanism of myosin {V}.
\newblock \emph{Proc. Natl. Acad. Sci. U.S.A.} 96:13726--13731.

\bibitem{Sellers10}
Sellers JR, Veigel C
\newblock (2010) Direct observation of the {myosin-Va} power stroke and its
  reversal.
\newblock \emph{Nat. Struct. Mol. Biol.} 17:590--U88.

\bibitem{Ortega11}
Ortega A, Amoros D, de~la Torre JG
\newblock (2011) Prediction of {hydrodynamic} and {other} {solution}
  {properties} of {rigid} {proteins} from {atomic-} and {residue-level} models.
\newblock \emph{Biophys. J.} 101:892--898.

\bibitem{Coureux04}
Coureux PD, Sweeney HL, Houdusse A
\newblock (2004) Three myosin {V} structures delineate essential features of
  chemo-mechanical transduction.
\newblock \emph{Embo J.} 23:4527--4537.

\bibitem{Thirum98}
Thirumalai D, Ha BY
\newblock (1998) in \emph{Theoretical and Mathematical Methods in Polymer
  Research}, ed{} Grosberg AY
\newblock ({Academic Press, New York}), pp 1--35.

\bibitem{delaCruz00}
{De La Cruz} EM, Wells AL, Sweeney HL, Ostap EM
\newblock ({2000}) {Actin and light chain isoform dependence of myosin V
  kinetics}.
\newblock \emph{{Biochemistry}} {39}:{14196--14202}.

\bibitem{Barclay98}
Barclay CJ
\newblock ({1998}) {Estimation of cross-bridge stiffness from maximum
  thermodynamic efficiency}.
\newblock \emph{{J. Muscle Res. Cell Motil.}} {19}:{855--864}.

\bibitem{Decostre05}
Decostre V, Bianco P, Lombardi V, Piazzesi G
\newblock (2005) Effect of temperature on the working stroke of muscle myosin.
\newblock \emph{Proc. Natl. Acad. Sci. U.S.A.} 102:13927--13932.

\bibitem{Lewalle08}
Lewalle A, Steffen W, Stevenson O, Ouyang Z, Sleep J
\newblock ({2008}) {Single-molecule measurement of the stiffness of the rigor
  myosin head}.
\newblock \emph{{Biophys. J.}} {94}:{2160--2169}.

\bibitem{Sun11}
Sun YJ, Goldman YE
\newblock (2011) Lever-arm {mechanics} of {processive} myosins.
\newblock \emph{Biophys. J.} 101:1--11.

\bibitem{Spink08}
Spink BJ, Sivaramakrishnan S, Lipfert J, Doniach S, Spudich JA
\newblock (2008) Long single alpha-helical tail domains bridge the gap between
  structure and function of myosin {VI}.
\newblock \emph{Nat. Struct. Mol. Biol.} 15:591--597.

\bibitem{Mukherjea09}
Mukherjea M, {et~al.}
\newblock (2009) Myosin {VI} {dimerization} {triggers} an {unfolding} of a
  {three-helix} {bundle} in {order} to {extend} {its} reach.
\newblock \emph{Mol. Cell} 35:305--315.

\bibitem{Zhang10}
Thirumalai D, Zhang ZC
\newblock (2010) Myosin {VI:} {How} {do} {charged} {tails} {exert} control?
\newblock \emph{Structure} 18:1393--1394.

\bibitem{vanKampen}
van Kampen NG
\newblock (2007) \emph{Stochastic Processes in Physics and Chemistry, Third
  Edition (North-Holland Personal Library)}
\newblock ({North Holland}).

\bibitem{Wilemski1974a}
Wilemski G, Fixman M
\newblock (1974) Diffusion-controlled intrachain reactions of polymers 1.
  theory.
\newblock \emph{J. Chem. Phys.} 60:866--877.

\bibitem{Kratky49}
Kratky O, Porod G
\newblock (1949) Rontgenuntersuchung geloster fadenmolekule.
\newblock \emph{Recueil Des Travaux Chimiques Des Pays-bas-journal Royal
  Netherlands Chem. Soc.} 68:1106--1122.

\bibitem{Saito67}
Saito N, Takahashi W, Yunoki Y
\newblock (1967) Statistical mechanical theory of stiff chains.
\newblock \emph{J. Phys. Soc. Japan} 22:219--226.

\bibitem{Hyeon06}
Hyeon C, Thirumalai D
\newblock (2006) Kinetics of interior loop formation in semiflexible chains.
\newblock \emph{J. Chem. Phys.} 124:104905.

\end{thebibliography}

\begin{thebibliography}{10}

\bibitem{vanKampen}
van Kampen NG
\newblock (2007) \emph{Stochastic Processes in Physics and Chemistry, Third
  Edition (North-Holland Personal Library)}
\newblock ({North Holland}).

\bibitem{Coureux04}
Coureux PD, Sweeney HL, Houdusse A
\newblock (2004) Three myosin {V} structures delineate essential features of
  chemo-mechanical transduction.
\newblock \emph{Embo J.} 23:4527--4537.

\bibitem{Ortega11}
Ortega A, Amoros D, de~la Torre JG
\newblock (2011) Prediction of {hydrodynamic} and {other} {solution}
  {properties} of {rigid} {proteins} from {atomic-} and {residue-level} models.
\newblock \emph{Biophys. J.} 101:892--898.

\bibitem{Guo95BP}
Guo Z, Thirumalai D
\newblock (1995) {Kinetics of protein folding: nucleation mechanism, time
  scales, and pathways}.
\newblock \emph{Biopolymers} 36:83--102.

\bibitem{Ermak78}
Ermak DL, McCammon JA
\newblock (1978) Brownian dynamics with hydrodynamic interactions.
\newblock \emph{J. Chem. Phys.} 69:1352--1360.

\bibitem{Rotne69}
Rotne J, Prager S
\newblock (1969) Variational treatment of hydrodynamic interaction in polymers.
\newblock \emph{J. Chem. Phys.} 50:4831--4837.

\bibitem{DoiEdwards}
Doi M, Edwards SF
\newblock (1988) \emph{The Theory of Polymer Dynamics}
\newblock ({Oxford University Press, USA}).

\bibitem{Wegener80}
Wegener WA
\newblock (1980) Hydrodynamic resistance and diffusion-coefficients of a freely
  hinged rod.
\newblock \emph{Biopolymers} 19:1899--1908.

\bibitem{Dunn07}
Dunn AR, Spudich JA
\newblock (2007) Dynamics of the unbound head during myosin {V} processive
  translocation.
\newblock \emph{Nat. Struct. Mol. Biol.} 14:246--248.

\bibitem{Kratky49}
Kratky O, Porod G
\newblock (1949) Rontgenuntersuchung geloster fadenmolekule.
\newblock \emph{Recueil Des Travaux Chimiques Des Pays-bas-journal Royal
  Netherlands Chem. Soc.} 68:1106--1122.

\bibitem{Saito67}
Saito N, Takahashi W, Yunoki Y
\newblock (1967) Statistical mechanical theory of stiff chains.
\newblock \emph{J. Phys. Soc. Japan} 22:219--226.

\bibitem{Thirum98}
Thirumalai D, Ha BY
\newblock (1998) in \emph{Theoretical and Mathematical Methods in Polymer
  Research}, ed{} Grosberg AY
\newblock ({Academic Press, New York}), pp 1--35.

\bibitem{Chiu94}
Chiu LYC, Moharerrzadeh M
\newblock (1994) Translational and rotational expansion of spherical gaussian
  wave-functions for multicenter molecular integrals.
\newblock \emph{J. Chem. Phys.} 101:449--458.

\bibitem{Cohen91}
Cohen A
\newblock (1991) A pade approximant to the inverse langevin function.
\newblock \emph{Rheol. Acta} 30:270--273.

\bibitem{Wang03}
Wang HY, Peskin CS, Elston TC
\newblock (2003) A robust numerical algorithm for studying biomolecular transport processes.
\newblock \emph{J. Theor. Biol.} 221:491--511.

\end{thebibliography}
\end{document}